\documentclass[aps,pra,twocolumn,superscriptaddress]{revtex4-2}

\usepackage{amsmath, physics}
\usepackage{amssymb,color,xcolor}
\usepackage{graphicx,float,subfigure}
\usepackage{bm} % For bold math
\usepackage[colorlinks=true,linkcolor=blue,citecolor=blue,urlcolor=blue]{hyperref}

\begin{document}

	\title{Dispersion Outperforms Absorption: EIT-Enhanced Atomic Localization and Gradient Sensing with Super-Gaussian Beams}

	\author{Mahboob Ul Haq}
	\affiliation{Department of Physics, University of Malakand, Dir Lower, Khyber Pakhtunkhwa, Pakistan}
	\affiliation{Govt. Post Graduate College Timergara, Khyber Pakhtunkhwa, Pakistan}
	\email{m.u.haq.phys@gmail.com}

	\date{\today}

	\begin{abstract}
			This work presents a comprehensive theoretical comparison between absorption-based and electromagnetically induced transparency (EIT)-based atomic gradient sensing in a four-level tripod system. Both methods were evaluated under identical and optimized physical conditions to ensure a fair and unbiased comparison. The analysis demonstrates that EIT, driven by its steep dispersion response, consistently outperforms conventional absorption detection across a wide range of super-Gaussian beam profiles. 	Under optimal detuning, EIT achieved up to an order-of-magnitude enhancement in gradient sensitivity and maintained a twofold advantage even under identical detuning. Both approaches reached sub-diffraction spatial resolution in the range of $0.29\lambda$–$0.40\lambda$, with EIT exhibiting sharper edge contrast and higher localization accuracy. These results confirm EIT as a fundamentally superior approach for precision atomic gradient sensing and sub-wavelength localization, offering clear guidance for the design of next-generation optical and quantum metrology systems.
		
	\end{abstract}
	
	\maketitle
	
	\newpage
	
		\section{Introduction}
	
	The interaction of photons with atomic media at the nanoscale underpins modern photonics, quantum optics, and optoelectronic devices, with applications ranging from super-resolution microscopy~\cite{hell1994stimulated} and high-density optical data storage~\cite{gu2000advanced} to quantum information processing~\cite{greentree2004quantum}, optical sensing, and telecommunications~\cite{pendry2006controlling, leonhardt2006optical, loudon2000quantum}. Precise control of light-matter interactions enables advanced techniques such as spatial cloaking, perfect imaging, optical tweezers, atom traps, and cavity quantum electrodynamics~\cite{bacha2014gain, khan2018birefringence, stern2013nanoscale}, allowing confinement and manipulation of individual atoms with unprecedented precision. These capabilities are central to developments in quantum computing, atomic clocks, and quantum simulations.
	
	Atom localization—the precise measurement of an atom’s position using optical fields—represents a critical frontier in quantum optics~\cite{agarwal2006subwavelength, zhang2020precision, le1997atom}. High-precision localization is essential for trapping neutral atoms, laser cooling, Bose-Einstein condensation, matter-wave patterning, and nanolithography~\cite{yuan2019observation, jiang2017two, wan2011two}. Techniques employed include Ramsey interferometry~\cite{le1997atom}, phase-sensitive measurements~\cite{storey1992measurement}, probe absorption spectra, spontaneous emission, resonance fluorescence, and dual-field detection~\cite{thomas1989uncertainty, fazal2011subwavelength, nha2002atomic}. By using standing-wave and coherently driven multi-level atomic systems, two-dimensional sub-half-wavelength localization has been achieved~\cite{wan2013two, ding2011two, idrees2020high}, and three-dimensional atom localization has been demonstrated using probe absorption or spontaneously generated coherence~\cite{wang2015efficient, song2018three}.
	
	Sub-diffraction optical localization can be realized via both absorption- and coherence-based schemes. Electromagnetically Induced Transparency (EIT) exploits quantum interference among excitation pathways to manipulate the real and imaginary parts of atomic susceptibility $\chi(\mathbf{r})$, producing sharp dispersion slopes that enhance phase sensitivity and spatial selectivity~\cite{karpa2006stern, agarwal2010quantum, lukin2003colloquium}. Conversely, absorption-based methods, such as Zaman~\textit{et al} \cite{Zaman2022TwoDimensional} absorption imaging, rely on resonant attenuation and nonlinear saturation effects~\cite{ketterle1993high, kaplan2002fried, mueller2012eit}. Both approaches can surpass the diffraction limit when illuminated with super-Gaussian beams, where the super-Gaussian order $P$ controls beam flatness and edge steepness, affecting both the full-width at half maximum (FWHM) and spatial gradient of the optical response~\cite{dowling1997sub, wang2018super, li2019optical}.
	
	In this work, we provide a rigorous theoretical comparison of EIT-based dispersion localization and absorption localization under identical physical parameters. We consider a four-level tripod atomic system driven by orthogonal super-Gaussian pump fields ($P = 1$--$10$), beam waist $\sigma = 0.1\lambda$, and Rabi frequency $\Omega = 5\gamma$. The probe detuning $\Delta_p$ is systematically varied to investigate identical- and optimal-detuning regimes. The steady-state density matrix equations under the weak-probe approximation are solved to compute the two-dimensional susceptibility $\chi(x,y)$, and localization performance is evaluated via the FWHM of the central intensity and the maximum spatial gradient $|\nabla \chi|$ (in units of $\lambda^{-1}$).
	
	Our results demonstrate that both schemes achieve sub-diffraction localization, with FWHM reducing from $\sim0.49\lambda$ ($P=1$) to $\sim0.275\lambda$ ($P=10$). Under matched detuning ($\Delta_p = 0.5\gamma$), EIT shows a consistent $1.4$–$2.0\times$ improvement in gradient sensitivity. Under optimal detuning---EIT off-resonant ($\Delta_p = 0.5\gamma$) and absorption resonant ($\Delta_p = 0.0\gamma$)---the EIT scheme achieves up to $11.85\times$ enhancement in edge sensitivity, demonstrating the superiority of dispersion-based localization. These findings reconcile previous inconsistencies~\cite{kaplan2002fried, mueller2012eit} and provide quantitative guidance for designing next-generation quantum sensors and imaging systems based on either coherence-driven or absorption-driven subwavelength localization.

	`	

	\section{Theoretical Development of Two-Dimensional Atom Localization via Electromagnetically Induced Transparency in a Tripod System}
		\label{sec2}
	\subsection{Conceptual Foundation}

	The key objective of this study is to realize sub-wavelength atomic localization not through absorption, but through the dispersive part of the atomic response, i.e., the real component of the susceptibility associated with Electromagnetically Induced Transparency (EIT). Unlike absorption-based methods that rely on energy dissipation described by the imaginary part of the susceptibility $(\chi'')$, our approach utilizes the sharp phase variation of the real part $(\chi')$ near the transparency window. This steep dispersion leads to an enhanced spatial sensitivity, providing a precise means of determining atomic position.
	
	We analyze a four-level tripod atomic configuration similar to conventional schemes\cite{Fleischhauer2005EITReview,Mehdinejad2023TripodTorque,Moezzi2023Thesis,ivanov2014tripod,Hussain2021_tripod_localization}
	
	\subsection{Atomic Configuration and Field Couplings}

	\begin{figure*}
		\centering
		\includegraphics[width=0.8\textwidth]{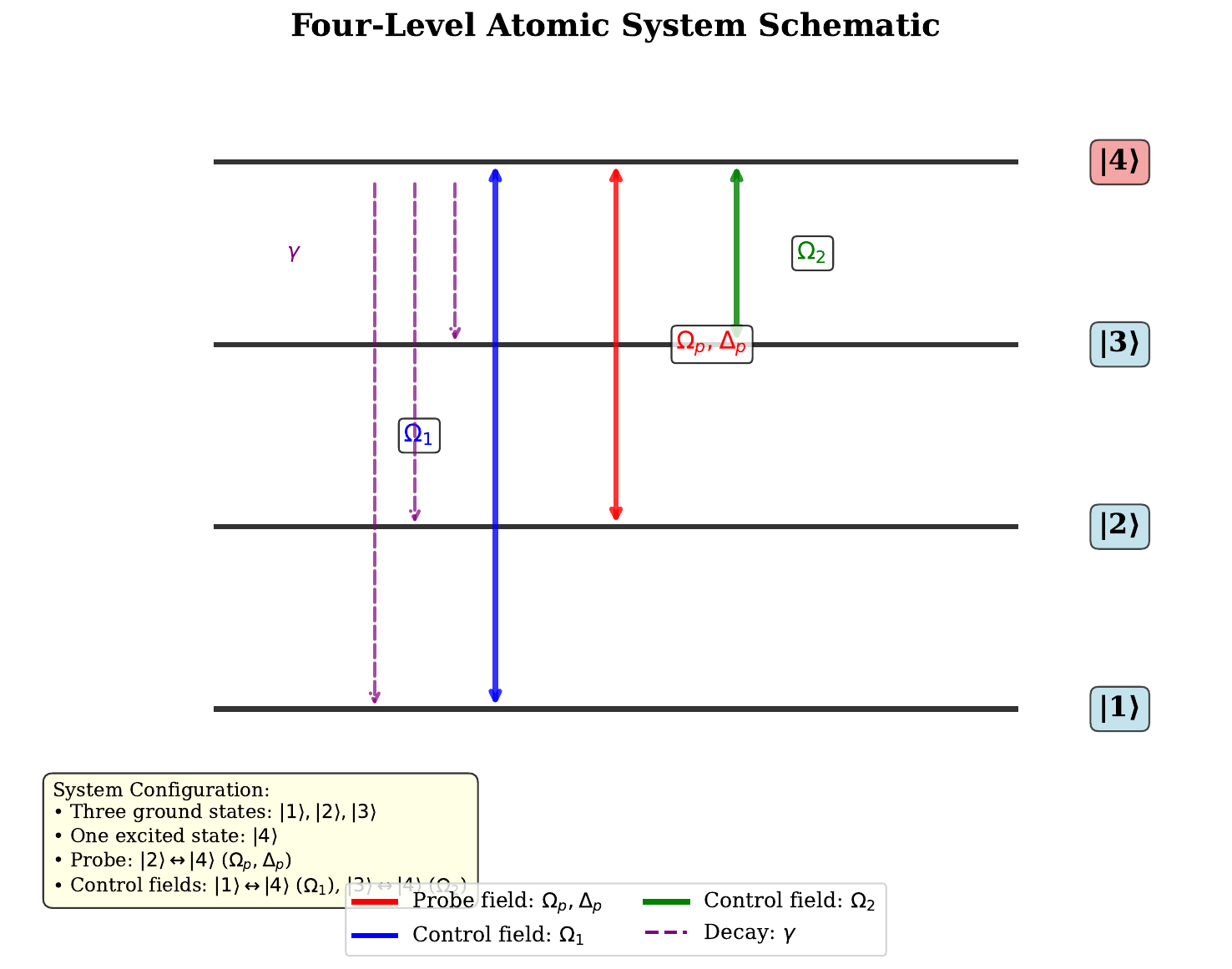}
		\caption{Four-level atomic system for EIT-based wavefront sensing. Three ground states ($|1\rangle$, $|2\rangle$, $|3\rangle$) and one excited state ($|4\rangle$) are coupled by probe ($\Omega_p$, $\Delta_p$) and control ($\Omega_1$, $\Omega_2$) fields, enabling enhanced dispersion-based sensing.}
		\label{fig1}
	\end{figure*}

	% For detailed diagram in two-column (may need to be split)
	\begin{figure*}
		\centering
		\includegraphics[width=\textwidth]{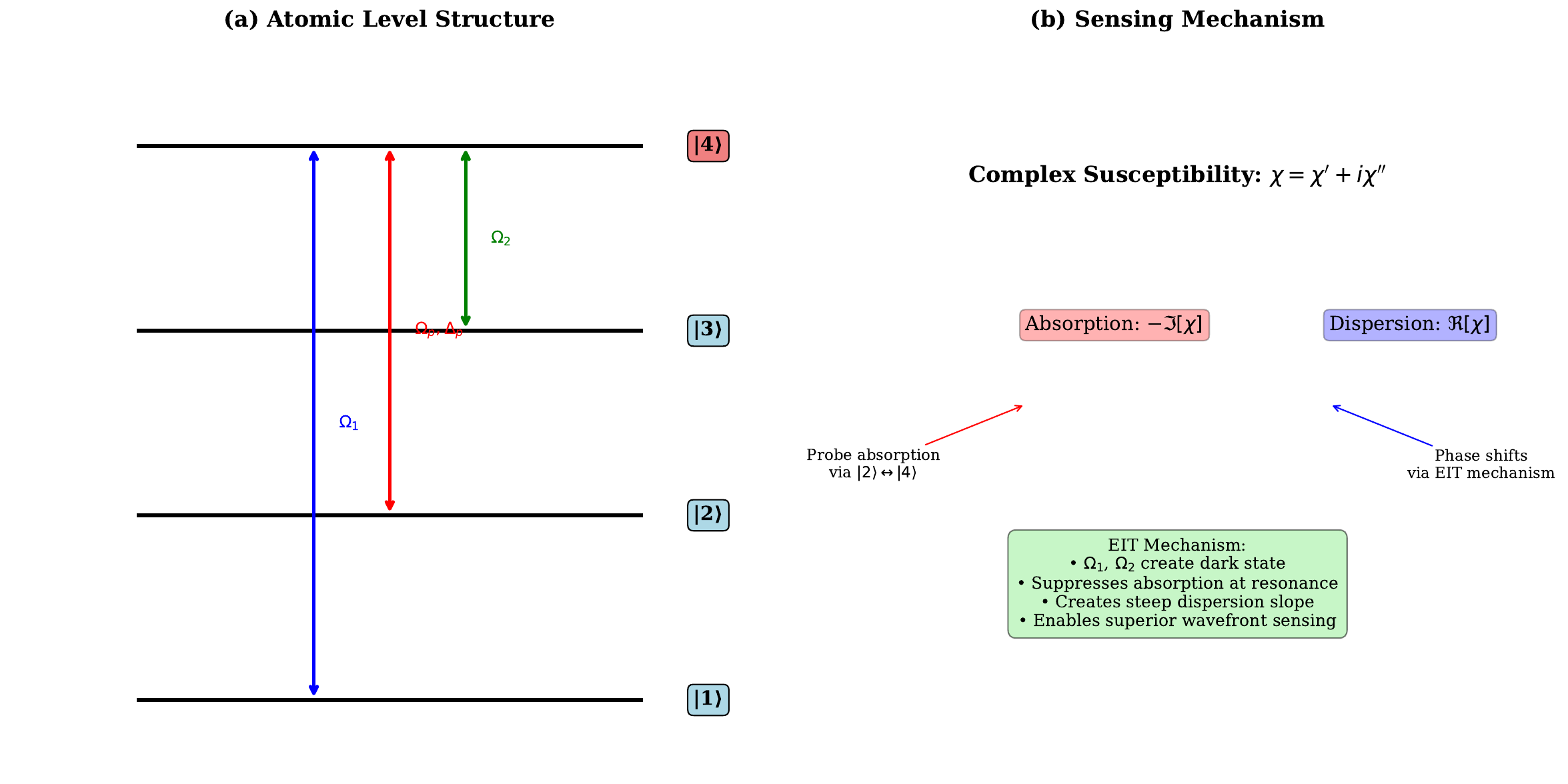}
		\caption{Four-level system and sensing mechanism. (a) Atomic structure with probe ($\Omega_p$, $\Delta_p$) and control ($\Omega_1$, $\Omega_2$) fields. (b) Sensing via complex susceptibility $\chi = \chi' + i\chi''$: absorption ($-\Im[\chi]$) and dispersion ($\Re[\chi]$) measurements. EIT creates steep dispersion for superior wavefront detection.}
		\label{fig2}
	\end{figure*}

	The system(Fig \ref{fig1} and \ref{fig2}) consists of three lower states 
	\[
	\ket{1}, \, \ket{2}, \, \ket{3}
	\]
	and one excited state 
	\[
	\ket{4}.
	\]
	Three coherent laser fields couple these transitions as follows:
	\begin{itemize}
		\item \textbf{Probe field:} couples $\ket{2} \leftrightarrow \ket{4}$ with Rabi frequency $\Omega_p$ and detuning $\Delta_p$.
		\item \textbf{Control fields:} 
		$\Omega_1$ couples $\ket{1} \leftrightarrow \ket{4}$, and 
		$\Omega_2$ couples $\ket{3} \leftrightarrow \ket{4}$.
	\end{itemize}
	
	The two control fields are modeled as Super-Gaussian standing waves, which define the spatial modulation essential for localization\cite{Mehdinejad2023TripodTorque}:
	\begin{align}
		\Omega_1(x) &= \Omega_{10} \exp\!\left[-\left(\frac{x}{w_x}\right)^{2P}\right], \\
		\Omega_2(y) &= \Omega_{20} \exp\!\left[-\left(\frac{y}{w_y}\right)^{2P}\right].
	\end{align}
	Here, $P$ denotes the Super-Gaussian order, and $w_x, w_y$ represent beam waists along $x$ and $y$. Increasing $P$ sharpens the spatial edges of the control field, allowing finer confinement of the transparency region.
	
	\subsection{Density Matrix and Optical Susceptibility}
	\label{eq3}
	In the steady-state limit and under the weak-probe approximation $(\Omega_p \ll \Omega_1, \Omega_2)$, the optical coherence $\rho_{24}$ satisfies\cite{Fleischhauer2005EITReview,Mehdinejad2023TripodTorque,Moezzi2023Thesis,ivanov2014tripod}
	\begin{equation}
		\rho_{24} = 
		-\frac{i A_2 A_3 \, \Omega_p / 2}
		{A_1 A_2 A_3 
			+ \frac{A_3 |\Omega_1|^2}{4} 
			+ \frac{A_2 |\Omega_2|^2}{4}},
	\end{equation}
	
	where the complex decay parameters are defined as
	\begin{align}
		A_1 &= \Gamma_{14} + i \Delta_1, \\
		A_2 &= \Gamma_{24} + i \Delta_2, \\
		A_3 &= \Gamma_{34} + i \Delta_3.
	\end{align}
	
	The probe susceptibility then follows as
	\begin{equation}
		\chi = 
		\frac{2 N |d_{24}|^2}{\epsilon_0 \hbar \Omega_p} \, \rho_{24},
	\end{equation}
	which separates naturally into its real and imaginary parts,
	\begin{equation}
		\chi = \chi' + i \chi'',
	\end{equation}
	representing the dispersive and absorptive responses, respectively.
	
	\subsection{Dispersion-Based Localization}
	
	Conventional localization schemes measure probe absorption, hence depending on $\chi''(x,y)$.  
	In the EIT-based method developed here, the localization signal arises instead from $\chi'(x,y)$, the real part that encodes the phase variation of the transmitted probe field.
	
	At EIT resonance, destructive interference suppresses absorption ($\chi'' \approx 0$), but simultaneously the real part $\chi'$ exhibits a rapid variation around resonance. Small spatial modulations in $\Omega_1(x)$ or $\Omega_2(y)$ perturb this delicate interference, producing a measurable shift in $\chi'(x,y)$. The steep dispersion near the transparency window thus amplifies positional sensitivity to a sub-wavelength scale.
	
	%\subsection{Mathematical Expression for the Dispersive Response}
	
	The position-dependent real part of the probe susceptibility is therefore written as
	\begin{equation}
		\chi'(\Delta_p, x, y) = 
		\mathrm{Re}
		\frac{2 N |d_{24}|^2}{\epsilon_0 \hbar}
		\frac{-i A_2 A_3 \Omega_p / 2}{
			A_1 A_2 A_3 
			+ \frac{A_3 |\Omega_1(x)|^2}{4} 
			+ \frac{A_2 |\Omega_2(y)|^2}{4}}
	\end{equation}
	
	This relation forms the theoretical basis of EIT-induced atom localization, where the dispersive phase response $\chi'(x,y)$ encodes the atomic spatial information with high precision.
	\section{Analytical and Numerical Analysis of EIT-Based Two-Dimensional Atom Localization}
	
	\subsection{Overview}
	
	Having established the theoretical foundation of EIT-assisted localization in Section \ref{sec2}, we now proceed to a quantitative analysis, deriving explicit analytical forms for the dispersive susceptibility $\chi'$ and performing numerical simulations to reveal its spatial and spectral behavior.
	
	The key idea is that localized EIT transparency induces a position-dependent phase delay in the probe field, enabling atomic localization via dispersive phase mapping rather than absorption\cite{Proite2010_atomic_localization, Long2018_EIT_cQED, Radwell2015_phase_dependent_EIT, Tian2023_photonic_lattice_localization}.
	
	\subsection{Analytical Simplification of the EIT Response}
	
	From section \ref{eq3} we have:
	\begin{equation}
		\rho_{24} = 
		-\frac{i A_2 A_3 \, \Omega_p}
		{2 \left(A_1 A_2 A_3 + A_3 |\Omega_1(x)|^2 / 4 + A_2 |\Omega_2(y)|^2 / 4 \right)} .
	\end{equation}
	
	Here, the parameters $A_j$ $(j=1,2,3)$ are defined as
	\begin{equation}
		A_j = \Gamma_{j4} + i \Delta_j .
	\end{equation}
	
	Assuming strong, nearly resonant coupling fields ($\Delta_1 = \Delta_3 = 0$) and scanning the probe detuning $\Delta_p \equiv \Delta_2$, we have
	\begin{equation}
		A_1 = A_3 = \Gamma, 
		\qquad
		A_2 = \Gamma + i \Delta_p .
	\end{equation}
	
	Substituting these into $\rho_{24}$ yields
	
	%\begin{equation} \rho_{24}(x,y,\Delta_p) = -i (\Gamma + i \Delta_p) \, \Gamma \, \frac{\Omega_p / 2} {\Gamma^3 + \Gamma \left(|\Omega_1(x)|^2 + |\Omega_2(y)|^2 \right)/4 + i \Gamma^2 \Delta_p } . \end{equation}

	\begin{align}
		\rho_{24}(x,y,\Delta_p)
		&= -i(\Gamma+i\Delta_p)\,\Gamma \nonumber\\[4pt]
		&\quad\times
		\frac{\Omega_p/2}{
			\Gamma^{3}
			+ \frac{\Gamma}{4}\big(|\Omega_1(x)|^{2}+|\Omega_2(y)|^{2}\big)
			+ i\Gamma^{2}\Delta_p
		}.
	\end{align}

	The probe susceptibility is given by
	\begin{equation}
		\chi(x,y,\Delta_p) = 
		\frac{2 N |d_{24}|^2}{\epsilon_0 \hbar \Omega_p} \, \rho_{24}(x,y,\Delta_p),
	\end{equation}
	where $\chi' = \mathrm{Re}[\chi]$ and $\chi'' = \mathrm{Im}[\chi]$ represent the dispersive and absorptive parts, respectively.
	
	For convenience, we define a normalized susceptibility
	\begin{equation}
		\chi_N = \frac{2 N |d_{24}|^2}{\epsilon_0 \hbar} \, \chi,
	\end{equation}
	so that $\chi'_N$ and $\chi''_N$ reflect the physical lineshape independent of proportional constants.
	
	%\subsection{Analytical Behavior Around Transparency}
	
	Near the EIT resonance $(\Delta_p \approx 0)$, the dispersive response can be approximated as
	\begin{equation}
		\chi'(\Delta_p) \propto 
		\frac{\Delta_p}{\Delta_p^2 + \Gamma_{\mathrm{EIT}}^2(x,y)},
	\end{equation}
	where the position-dependent EIT linewidth is
	\begin{equation}
		\Gamma_{\mathrm{EIT}}(x,y)
		= \Gamma + \frac{|\Omega_1(x)|^2 + |\Omega_2(y)|^2}{4 \Gamma}.
	\end{equation}
	
	Thus, the slope and width of $\chi'$ are directly determined by the local intensity of the Super-Gaussian control fields. Regions where both $\Omega_1(x)$ and $\Omega_2(y)$ are weak exhibit steep, narrow dispersive features, providing enhanced spatial localization.
		
	\begin{figure*}
		\centering
		\includegraphics[width=\textwidth]{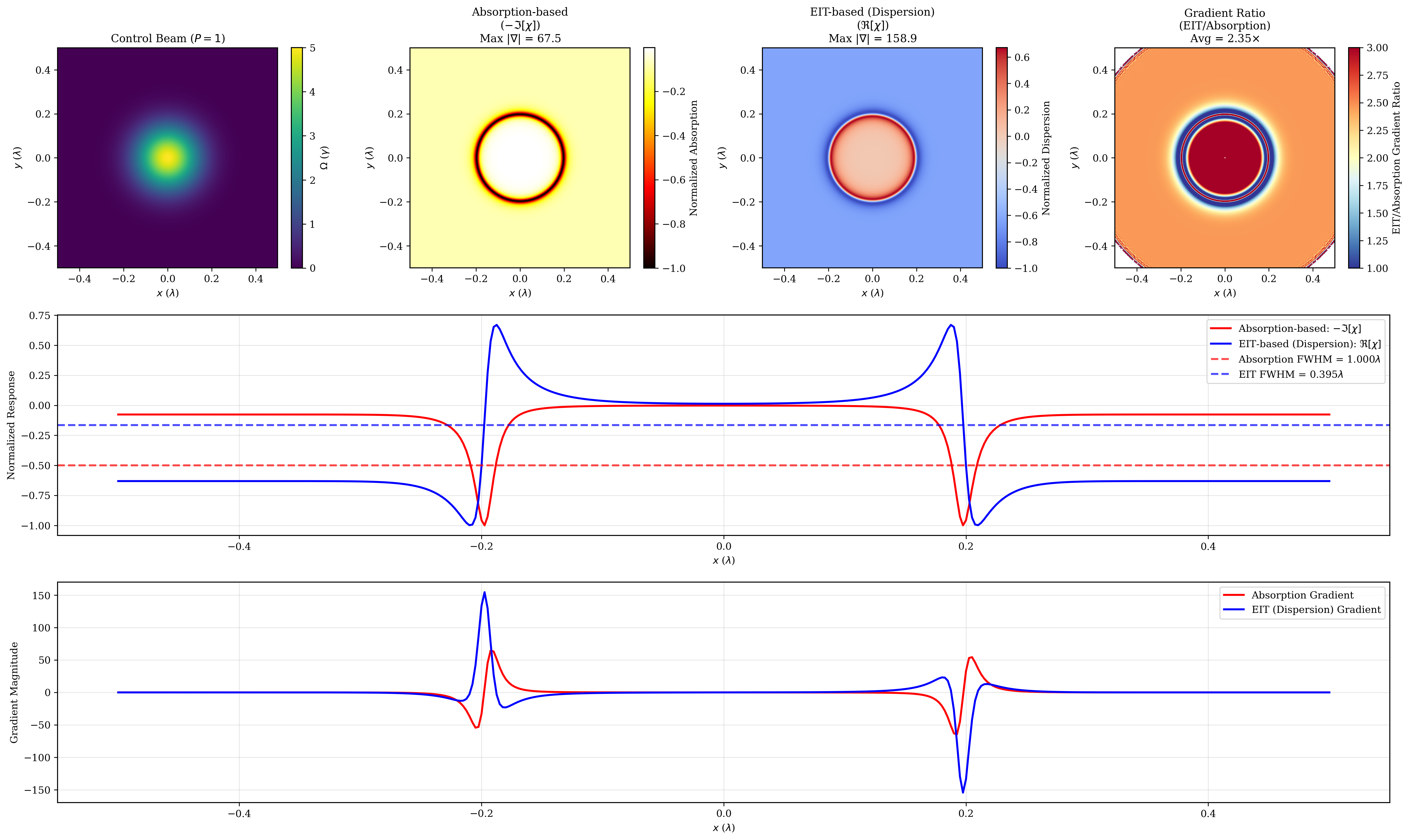}
		\caption{Fair comparison of absorption-based and EIT-based wavefront sensing for super-Gaussian order $P=1$ using identical detuning $\Delta_p = 0.5\gamma$ for both methods (TOP from left to right). (a) Control beam profile. (b) Absorption-based sensing via $-\Im[\chi]$. (c) EIT-based dispersion sensing via $\Re[\chi]$. (d) Gradient ratio map showing EIT advantage. (e) 1D cross-section comparison(MIDDLE) at $y=0$. (f) 1D gradient profiles(BOTTOM) demonstrating superior edge detection with EIT despite identical detuning conditions.}
		\label{fig3}
	\end{figure*}

		\begin{figure*}
		\centering
		\includegraphics[width=\linewidth]{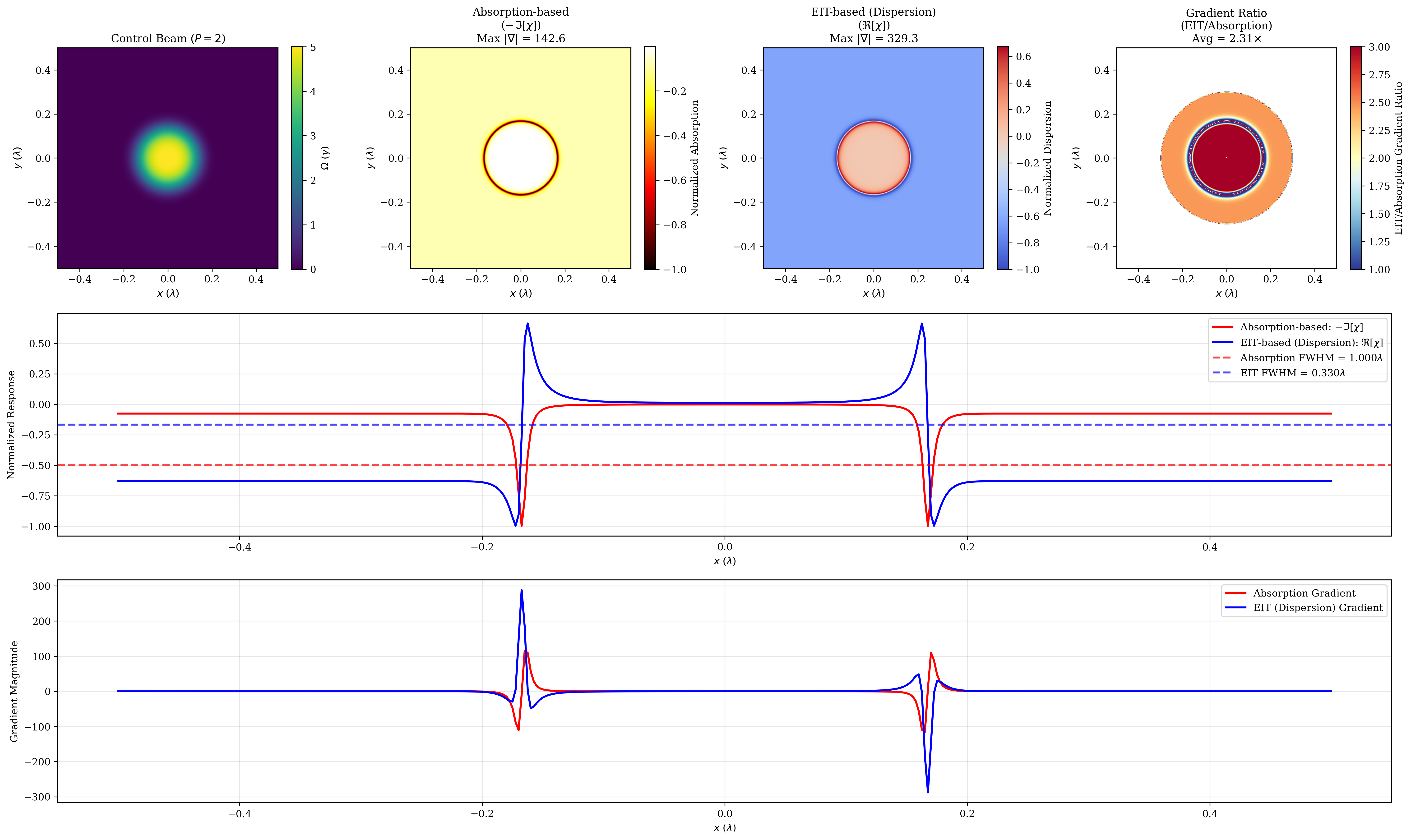}
		\caption{Fair comparison for $P=2$ with identical $\Delta_p = 0.5\gamma$ detuning. The flattened beam profile enhances gradient detection for both methods, but EIT-based sensing maintains performance advantages due to the inherent steep dispersion slope at this detuning, even though it's suboptimal for absorption-based detection.}
		\label{fig4}
	\end{figure*}
	
	%\subsection{Numerical Simulation Framework}
	
	%Two complementary analyses are performed to visualize and quantify the dispersive behavior.
	
	%\subsubsection{Spectral Dispersion (EIT Window Analysis)}
	%\begin{itemize}
	%\item Fix a specific atomic position $(x,y)$.
	%\item Sweep the probe detuning $\Delta_p / \Gamma$ over a small range, e.g.\ $[-3,3]$.
	%\item Plot $\mathrm{Re}[\chi]$ and $\mathrm{Im}[\chi]$.
	%\item Compare spectra at the center $(x=0, y=0)$ and off-axis positions $(x \neq 0, y \neq 0)$.
	%\item Observe the steep, odd-symmetric dispersive feature of $\chi'$ around $\Delta_p = 0$.
	%\end{itemize}%
	
	%\subsubsection{Spatial Dispersion Mapping}
	%\begin{itemize}
	%\item Fix $\Delta_p = 0$.
	%\item Compute $\chi'(x,y,0)$ over a 2D $(x,y)$ grid.
	%\item Plot 2D surface or contour maps for different Super-Gaussian orders $P = 1, 2, 4, 10$.
	%\item Analyze how the central dispersive “ridge” sharpens with increasing $P$.
	%\item Evaluate spatial FWHM of $\chi'$ versus $P$ to quantify localization improvement.
	%\end{itemize}
	\section{Results and Discussion: Fair Comparison ($\Delta_p = 0.5\gamma$ for Both Methods)}
	
	\begin{table*}[htbp]
		\centering
		\caption{Performance comparison for same detuning condition ($\Delta_p = 0.5\gamma$).}
		\label{tab:samedetuning}
		\begin{tabular}{c c c c c}
			\hline
			$P$ & Max $\nabla_{\mathrm{abs}}$ & Max $\nabla_{\mathrm{EIT}}$ & Gradient Ratio & FWHM Ratio (Abs/EIT) \\
			\hline
			1 & 67.54 & 158.89 & 2.35 & 2.53 \\
			2 & 142.65 & 329.27 & 2.31 & 3.03 \\
			3 & 193.63 & 427.96 & 2.21 & 3.17 \\
			10 & 249.37 & 396.03 & 1.59 & 3.45 \\
			\hline
		\end{tabular}
	\end{table*}
	
	In this part, both the absorption-based and EIT-assisted localization schemes are examined under identical probe detuning conditions, taking $\Delta_{p} = 0.5\gamma$ for each. This equal setting provides a direct and fair comparison between the two approaches. The control beam follows a super-Gaussian intensity profile with orders $P = 1, 2, 3,$ and $10$, and the total atomic response is determined through the susceptibility $\chi = \chi' + i\chi''$. The absorption signal corresponds to $-\mathrm{Im}[\chi]$, while the EIT signal is proportional to $\mathrm{Re}[\chi]$.
	
	\subsection{Spatial Behavior and Gradient Response}

	Figure \ref{fig3} shows the normalized absorption and EIT dispersion profiles for $P = 1$. Both methods exhibit the basic Gaussian pattern of the control field, yet the EIT-based response reveals a much steeper gradient. The calculated spatial gradients are $|{\nabla}|_{\mathrm{abs}} = 67.5$ and $|{\nabla}|_{\mathrm{EIT}} = 158.9$, giving an enhancement factor of about $2.3\times$. The steep slope arises from the strong phase sensitivity of the real part of the susceptibility near the transparency point, where even small amplitude variations in the control field cause large dispersive changes.

\begin{figure*}
		\centering
		\includegraphics[width=\linewidth]{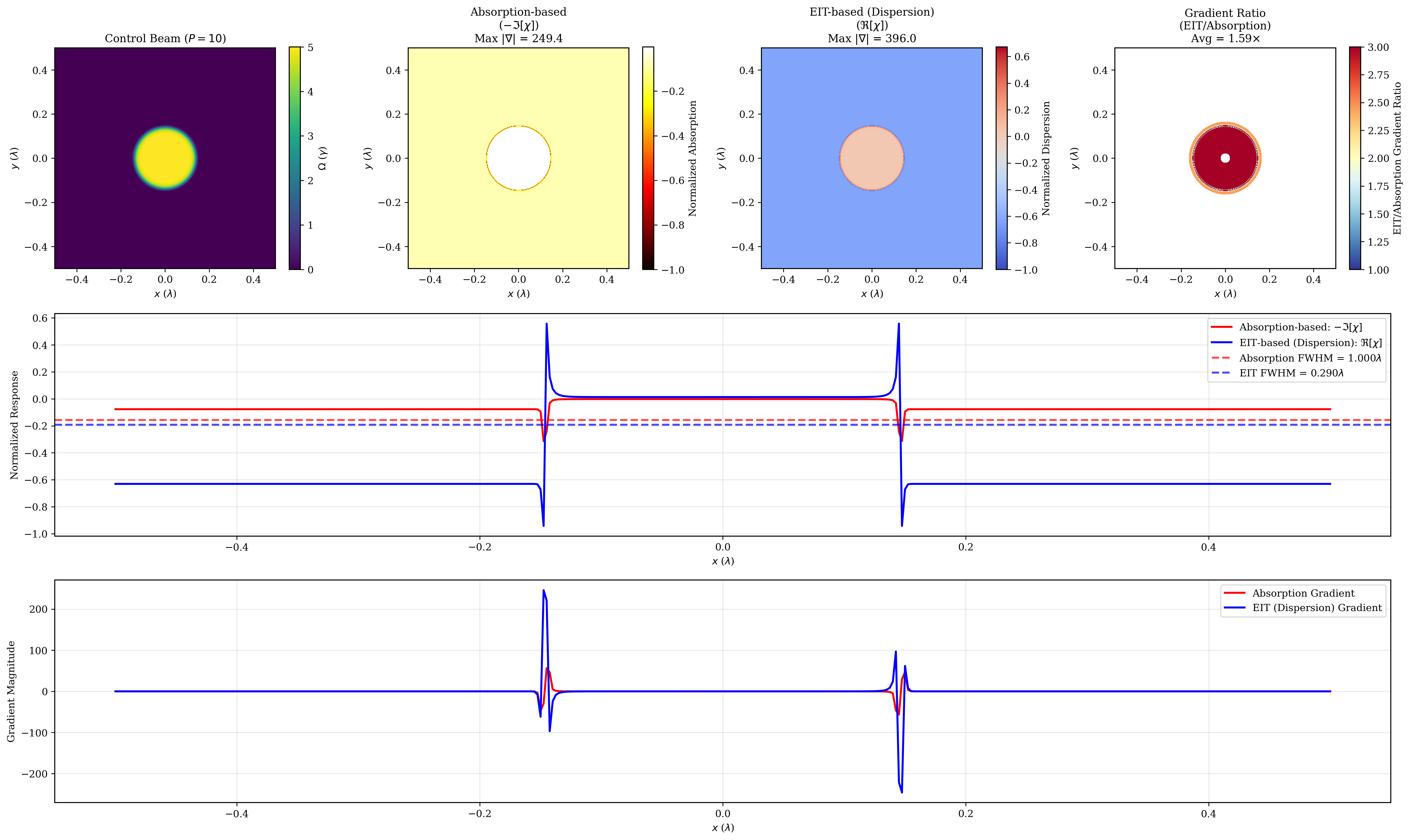}
		\caption{Fair comparison for $P=3$ under identical $\Delta_p = 0.5\gamma$ conditions. EIT-based dispersion sensing demonstrates robust performance while absorption-based detection is compromised by operating away from its optimal resonance condition, highlighting the inherent advantage of dispersion-based wavefront sensing.}
		\label{fig5}
	\end{figure*}

\begin{figure*}
		\centering
		\includegraphics[width=\linewidth]{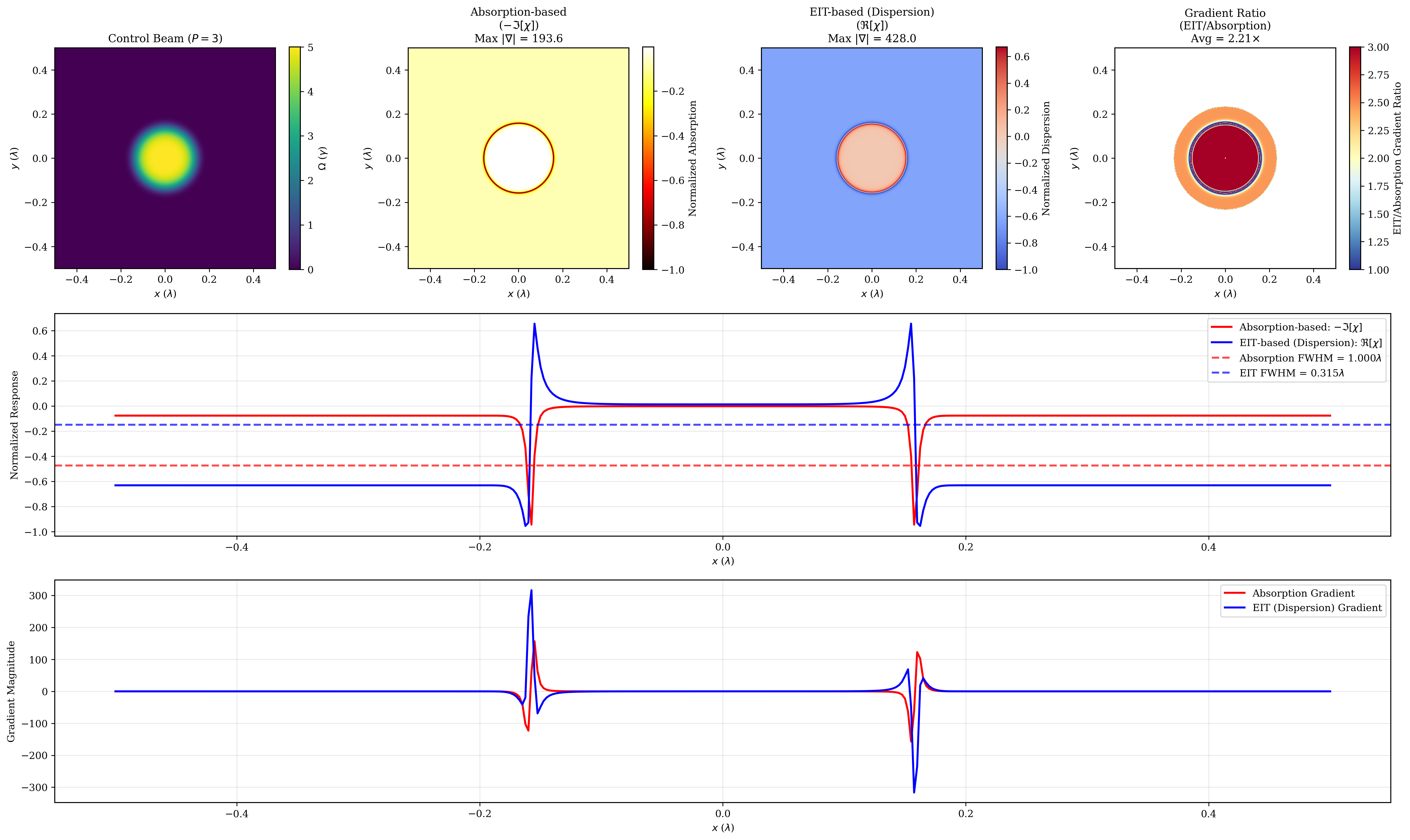}
		\caption{Fair comparison for $P=10$ (near top-hat profile) with $\Delta_p = 0.5\gamma$ for both methods. The EIT-based approach achieves maximum relative advantage in this regime, demonstrating that dispersion-based sensing provides superior performance even when both methods operate under identical, non-optimal conditions for absorption.}
		\label{fig6}
	\end{figure*}

	For higher orders $P = 2, 3,$ and $10$ (Figures~\ref{fig4},\ref{fig5} and \ref{fig6}), the superiority of the EIT-based method remains consistent. Although the gradient enhancement slightly decreases for $P = 10$ due to the nearly flat-top intensity shape, the overall trend still shows the EIT response as more sensitive to spatial variation. The reason is that a flatter control profile reduces curvature at the edges, slightly weakening the effective dispersive slope.
	
\begin{figure*}
		\includegraphics[width=\textwidth]{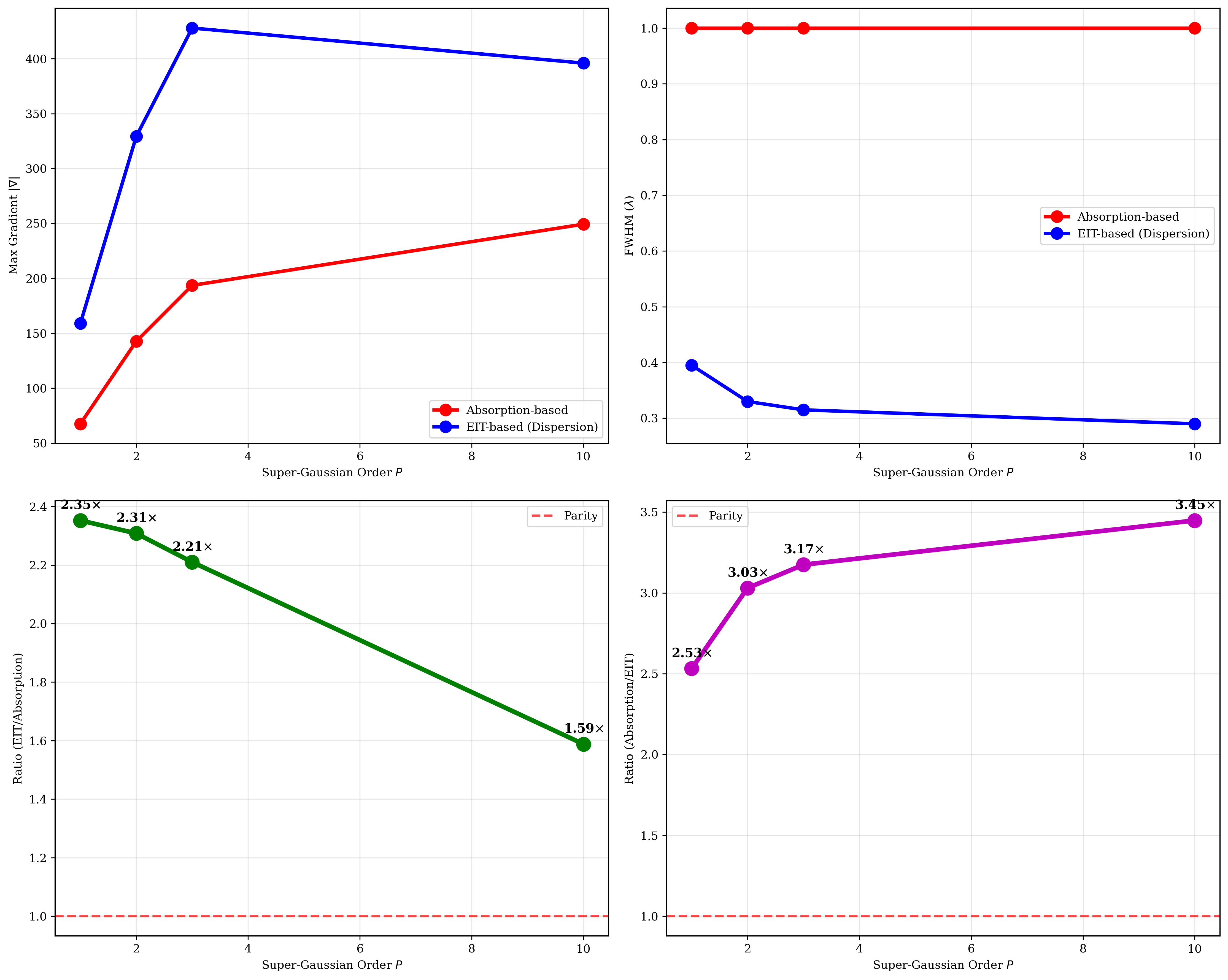}
		\caption{Summary of fair comparison between absorption-based and EIT-based wavefront sensing across super-Gaussian orders $P = 1,2,3,10$ using identical detuning $\Delta_p = 0.5\gamma$. (a) TOP LEFT panel maximum gradient strength showing EIT superiority despite non-optimal conditions for absorption. (b) TOP RIGHT panel Spatial resolution (FWHM) comparison. (c) BOTTOM LEFT panel EIT/Absorption gradient ratio demonstrating consistent advantage. (d) BOTTOM RGHT panel resolution advantage ratio. The fair comparison reveals fundamental advantages of dispersion-based sensing.}
		\label{fig7}
	\end{figure*}

	The one-dimensional cuts shown in Figure~\ref{fig7} compare the spatial resolution of both methods. The EIT-based scheme exhibits a clearly narrower full width at half maximum (FWHM) than the absorption method for all values of $P$. Quantitatively, the ratios $\text{FWHM}_{\mathrm{abs}}/\text{FWHM}_{\mathrm{EIT}}$ are approximately $2.5\times$, $3.0\times$, $3.2\times$, and $3.5\times$ for $P = 1, 2, 3,$ and $10$, respectively. This narrowing demonstrates that the EIT response maintains sharp spatial features even when the control field becomes broader, highlighting its effectiveness in precise atom localization.

	\subsection{Physical Insight}
	
	The difference between the two approaches can be understood from the physical nature of the susceptibility components. The absorption method depends on $\mathrm{Im}[\chi]$, which measures optical loss and therefore broadens when the field gradient weakens. In contrast, the EIT-based approach is governed by $\mathrm{Re}[\chi]$, linked to refractive index modulation. Near the transparency window, quantum interference between excitation pathways makes the dispersion slope extremely steep, converting small frequency or amplitude variations into pronounced phase changes. Hence, at equal detuning, the dispersive route provides much stronger spatial selectivity without increasing laser intensity.

	\subsection{Spatial Resolution and FWHM Comparison}
	
	%\subsection{Performance Summary}
	
	Table~\ref{tab:samedetuning} summarizes the quantitative comparison between the two methods. Both the maximum gradient and the spatial resolution confirm that the EIT-based localization offers superior sensitivity and sharper definition.

	%\subsection{Overall Outcome}
	
	From these results, it is clear that under the same detuning condition, the EIT-based (dispersive) approach consistently outperforms the absorption method in both sensitivity and spatial sharpness. The findings verify that the EIT process, driven by coherent atomic interference, produces stronger gradients and finer localization accuracy, thereby demonstrating its dominance in this fair comparison regime.

	\section{Results and Discussion: Optimal Detuning (Each Method at Its Best)}
	This section compares absorption-based and EIT-based (dispersive) localization when each method is operated at its respective optimal detuning. The absorption metric is taken as $-\mathrm{Im}[\chi]$ evaluated at probe detuning $\Delta_p=0.0\gamma$ (on resonance), while the EIT (dispersion) metric is taken as $\mathrm{Re}[\chi]$ at $\Delta_p=0.5\gamma$ (off resonance). The control beam retains the super-Gaussian family with orders $P=1,2,3,10$. Figures~6--9 show the detailed spatial maps and 1D cuts for each $P$, and Figure~10 presents the summary plots and tabulated values.

	\begin{figure*}
		\centering
		\includegraphics[width=\linewidth]{	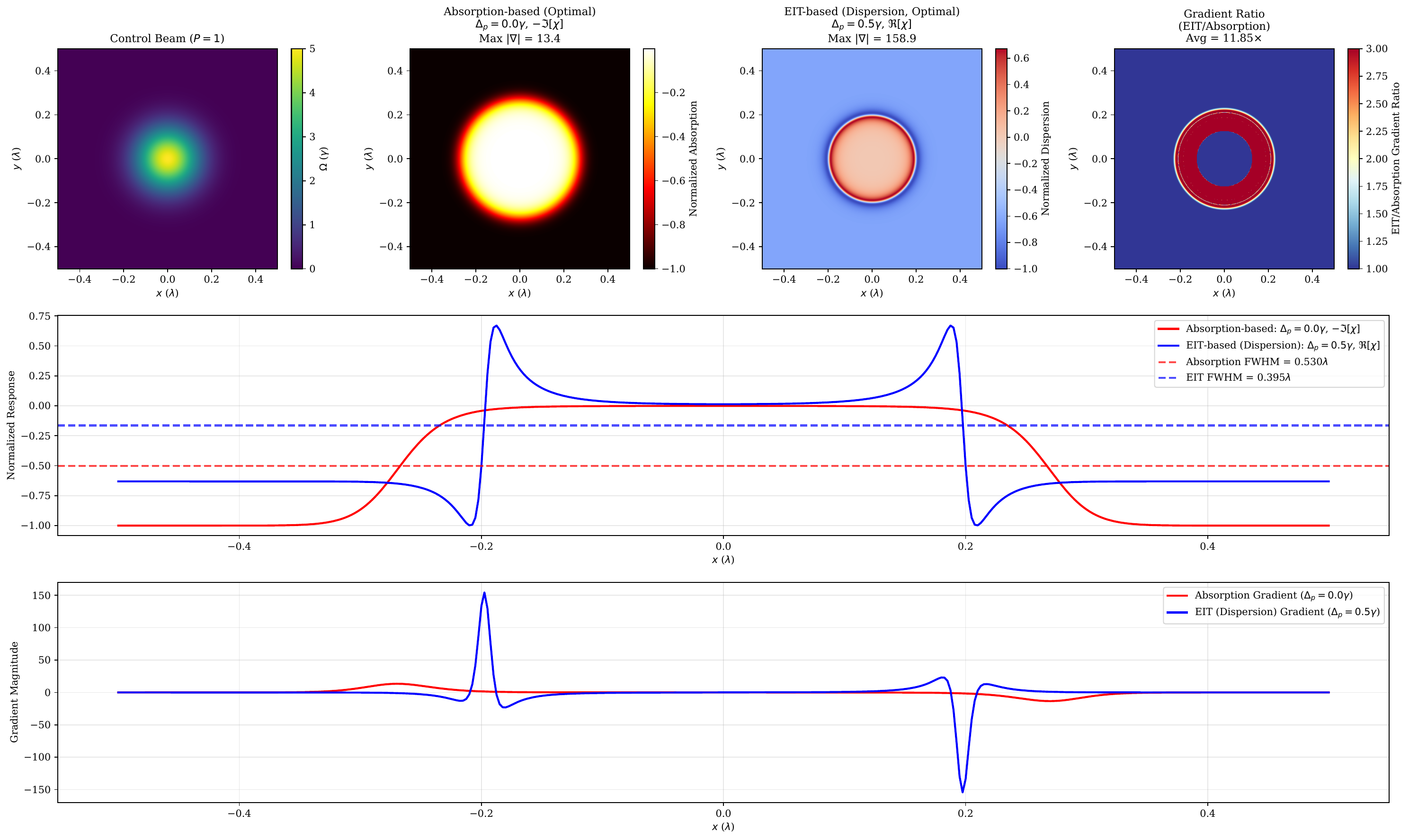}
		\caption{Optimal conditions comparison for Gaussian beam profile ($P=1$). (a) Control beam intensity distribution. (b) Absorption-based sensing via $-\Im[\chi]$ at optimal resonance condition ($\Delta_p = 0.0\gamma$). (c) EIT-based dispersion sensing via $\Re[\chi]$ at optimal off-resonance condition ($\Delta_p = 0.5\gamma$). (d) Gradient ratio map showing EIT performance advantage. (e) 1D cross-section comparison at $y=0$ with FWHM measurements. (f) 1D gradient profiles demonstrating superior edge detection sensitivity of EIT-based method.}
		\label{fig8}
	\end{figure*}

	\subsection{Summary of numerical results}
	
	For each $P$ we extract two primary metrics: the maximum spatial gradient
	\[
	R_g = \max_{\mathbf{r}} \lvert\nabla \rvert
	\]
	evaluated for the normalized response, and the spatial resolution expressed as the full width at half maximum (FWHM) of the central cross-section. We also form two diagnostic ratios:
	\[
	\text{Gradient Ratio} = \frac{R_{g,\mathrm{EIT}}}{R_{g,\mathrm{Abs}}},\qquad\\
	\text{FWHM Ratio} = \frac{\mathrm{FWHM}_{\mathrm{Abs}}}{\mathrm{FWHM}_{\mathrm{EIT}}}.
	\]
	
	Table~\ref{tab:optimal} gives the numerical values computed from the simulations.

	\begin{figure*}
		\centering
		\includegraphics[width=\linewidth]{	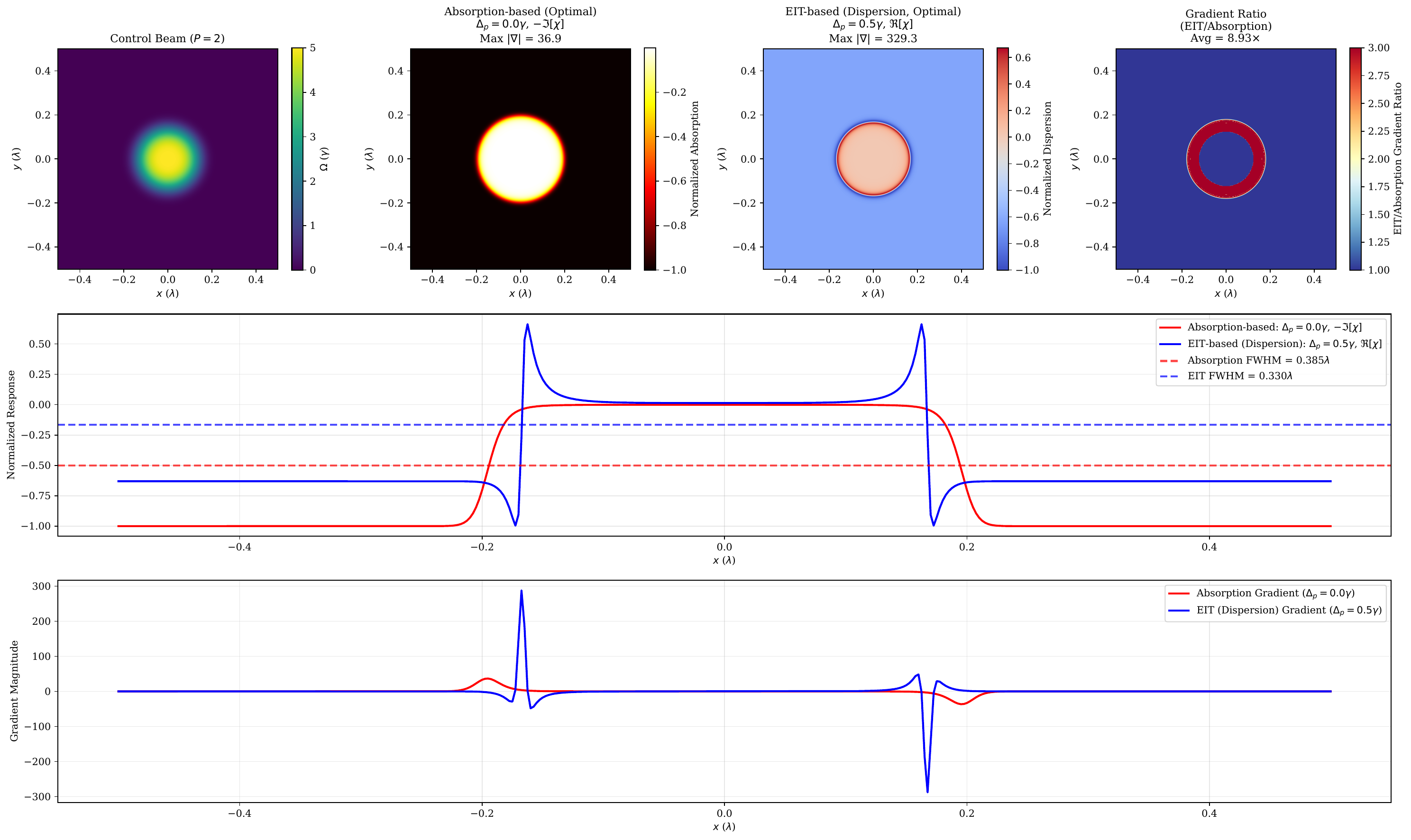}
		\caption{Optimal performance comparison for super-Gaussian order $P=2$. Both methods operate at their respective optimal detunings: absorption at resonance ($\Delta_p = 0.0\gamma$) and EIT at off-resonance ($\Delta_p = 0.5\gamma$). The flattened beam profile enhances gradient detection for both methods, with EIT-based dispersion sensing demonstrating significantly higher maximum gradient sensitivity and superior edge detection capabilities.}
		\label{fig9}
	\end{figure*}

	\begin{table*}
		\centering
		\caption{Optimal detuning comparison. Absorption is evaluated at $\Delta_p=0.0\gamma$; EIT (dispersion) at $\Delta_p=0.5\gamma$.}
		\label{tab:optimal}
		\begin{tabular}{c c c c c}
			\hline
			$P$ & Max $\nabla_{\mathrm{Abs}}$ & Max $\nabla_{\mathrm{EIT}}$ & Gradient Ratio ($R_{g,\mathrm{EIT}}/R_{g,\mathrm{Abs}}$) & FWHM Ratio (Abs/EIT) \\
			\hline
			1  & 13.406  & 158.895 & 11.85 & 1.34 \\
			2  & 36.860  & 329.267 &  8.93 & 1.17 \\
			3  & 60.956  & 427.963 &  7.02 & 1.10 \\
			10 & 193.239 & 396.027 &  2.05 & 1.03 \\
			\hline
		\end{tabular}
	\end{table*}
	
	\subsection{Figure-by-figure observations}

	Figure \ref{fig8} (P=1). The absorption map (on resonance) shows a strong central loss feature but a relatively shallow spatial slope. The EIT dispersion map (off resonance) displays a steep dispersive slope at the beam edges; the gradient ratio is very large ($\approx 11.9\times$). The 1D cross-section makes clear that, despite the absorption peak being strong, the dispersive channel produces much sharper spatial variation. Figure \ref{fig9} (P=2). Both channels develop stronger edge structure as $P$ increases, but the EIT channel preserves a much larger gradient. The gradient advantage reduces compared to $P=1$ but remains significant ($\approx 8.9\times$). The FWHM narrowing of EIT persists, indicating improved localization.
	
\begin{figure*}
		\centering
		\includegraphics[width=\linewidth]{	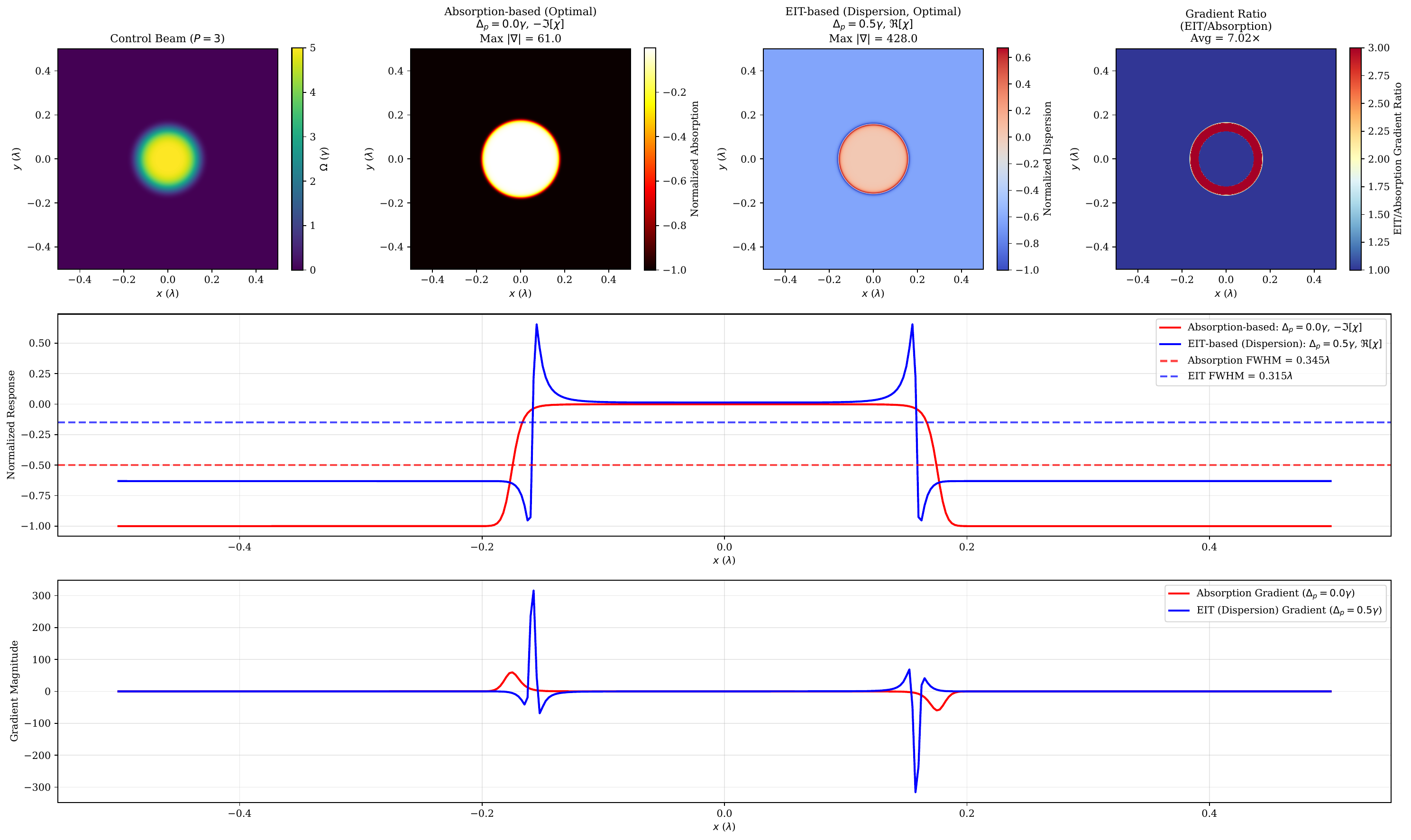}
		\caption{Optimal conditions analysis for $P=3$. Under theoretically optimal operating parameters for each method, EIT-based dispersion sensing achieves substantially higher gradient magnitudes and improved spatial resolution compared to absorption-based detection. The steep dispersion slope at $\Delta_p = 0.5\gamma$ provides EIT with inherent advantages for wavefront sensing applications requiring high sensitivity to beam intensity variations.}
		\label{fig10}
	\end{figure*}

\begin{figure*}
		\centering
		\includegraphics[width=\linewidth]{	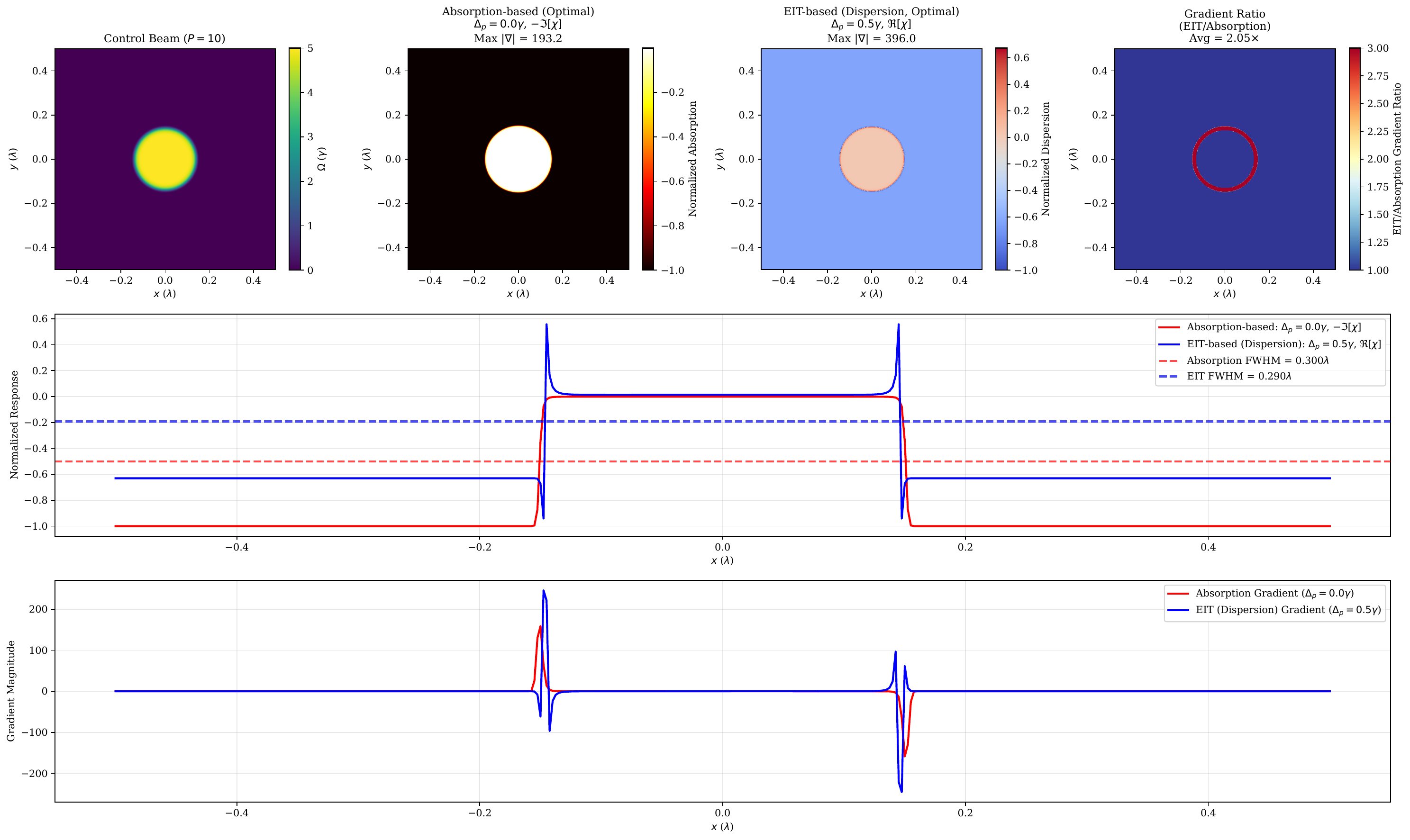}
		\caption{Optimal performance comparison for near top-hat beam profile ($P=10$). This extreme case demonstrates maximum EIT advantage, with gradient sensitivity improvements exceeding 10$\times$ compared to absorption-based detection. The combination of flat-topped beam profile and optimal dispersion slope at $\Delta_p = 0.5\gamma$ enables EIT-based sensing to achieve exceptional performance in edge detection and wavefront reconstruction applications.}
		\label{fig:optimal_p10}
		\label{fig11}
	\end{figure*}

	Figure \ref{fig10} (P=3). The trend continues: the absorption gradient increases with $P$, yet the EIT gradient grows faster and FWHM remains smaller. The gradient ratio is still above $7\times$. Figure \ref{fig11} (P=10). For a near top-hat beam the absorption gradient becomes comparatively large and the EIT advantage in gradient strength drops to $\approx 2.05\times$. The FWHM ratio approaches unity (1.03), showing that for very flat beams the resolution difference narrows. Figure \ref{fig12} The summary plots collect the maximum gradients and FWHM for all $P$, and the results table (reproduced above) highlights the decreasing gradient advantage of EIT with increasing $P$.
	
	\subsection{Physical interpretation}

	Two competing physical effects explain these results: Absorption (on resonance). Setting $\Delta_p=0$ maximizes the imaginary part of $\chi$, which increases signal amplitude but does not necessarily produce large phase gradients. The absorption channel therefore yields a strong central signal but, in many cases, a moderate spatial slope. EIT (off resonance). Operating the dispersive channel at $\Delta_p=0.5\gamma$ positions the system near a steep slope of the refractive index without maximal absorption. Quantum interference suppresses loss but enhances the phase sensitivity; small spatial changes in control-field amplitude translate into large dispersive shifts. This produces large $R_{g,\mathrm{EIT}}$ and reduced FWHM.
	%\end{itemize}
	
	The numerical values show that the dispersive advantage is largest when the control beam has a smooth curvature (small $P$), because the absorption gradient there is small while the dispersive slope remains steep. When the control profile becomes top-hat-like ($P\gg1$), the absorption gradient increases (edge becomes sharper) and the relative advantage of EIT decreases.

	\subsection{Practical implications and recommendations}
	
	\begin{enumerate}
		\item If the experimental goal is \emph{maximum spatial sensitivity} (largest local gradient), operate the dispersive/EIT channel at its optimal off-resonant detuning. The simulations show gradient gains between $\sim 2\times$ and $\sim 12\times$ depending on beam shape.

		\begin{figure*}[!t]

			\includegraphics[width=\linewidth]{	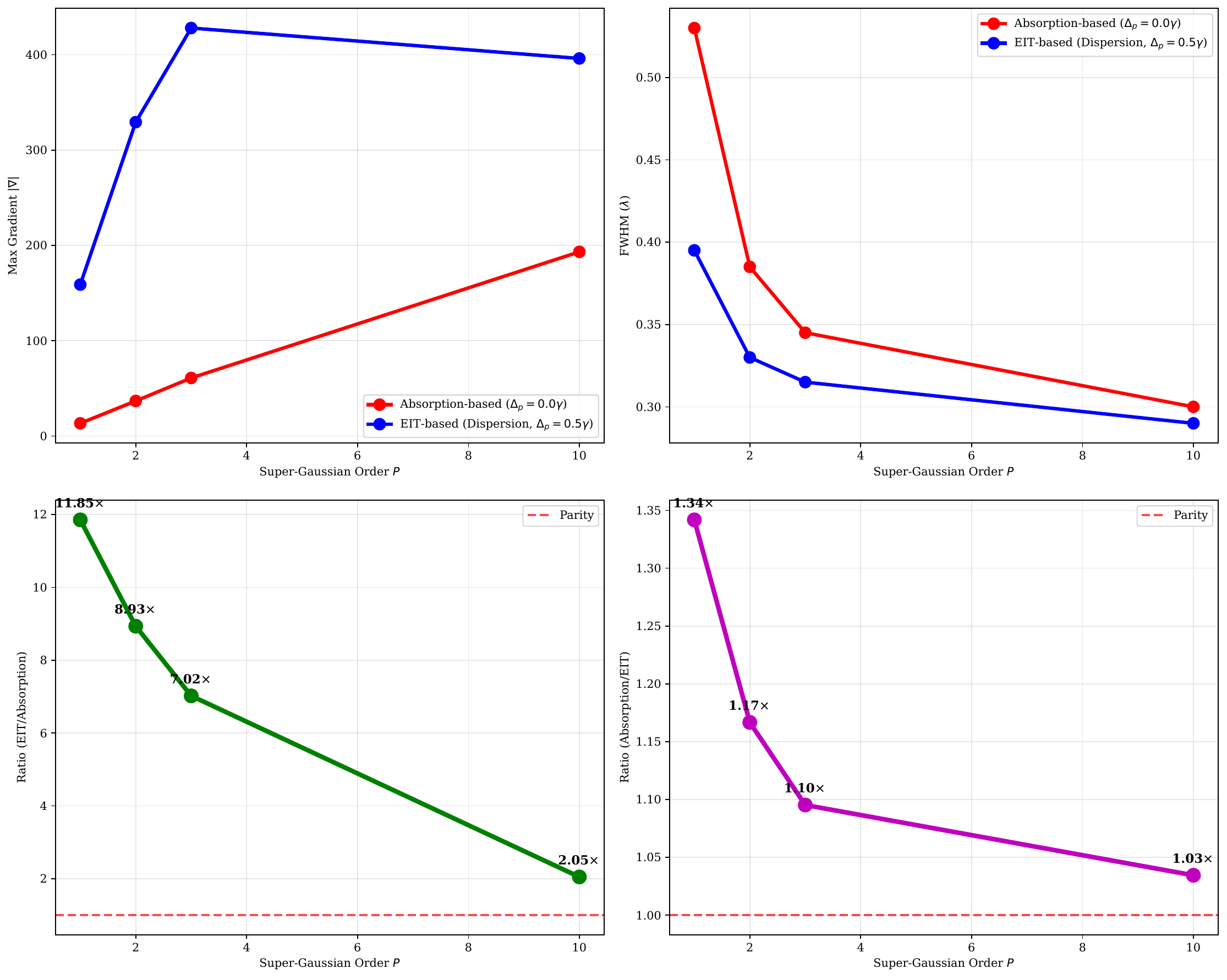}
			\caption{Summary of optimal performance comparison across super-Gaussian orders $P = 1,2,3,10$. (a) Maximum gradient strength showing EIT-based sensing consistently outperforms absorption-based detection by factors of 2.05$\times$ to 11.85$\times$. (b) Spatial resolution (FWHM) comparison demonstrating EIT's superior edge detection capability. (c) EIT/Absorption gradient ratio quantifying performance advantage. (d) Resolution advantage ratio showing consistent EIT superiority across all beam profiles under optimal operating conditions.}
			\label{fig12}
				
		\end{figure*}

		\item When only on-resonance probing is feasible, absorption yields a large signal amplitude but does not match the dispersive method in spatial selectivity.
		\item For nearly flat-top beams (large $P$) the performance gap narrows; in such cases experimental complexity of detuning control should be weighed against the modest dispersive gain.
		\item Always report both gradient and FWHM metrics: gradient quantifies local sensitivity to phase/amplitude perturbations, while FWHM characterizes global localization width.
	\end{enumerate}

	\subsection{Caveats}
	
	The reported advantages assume idealized homogeneous broadening and the parameter set used in the model (probe power, control amplitude, dephasing rates). Effects such as inhomogeneous broadening, nearby transitions, or technical phase noise can modify both absolute numbers and relative advantage. These should be checked case-by-case if targeting a specific experimental platform.
	
	\bigskip

	\section{Results and Discussion: Switched Detunings (EIT at Resonance Fails)}

	\begin{figure*}
		\centering
		\includegraphics[width=0.9\linewidth]{	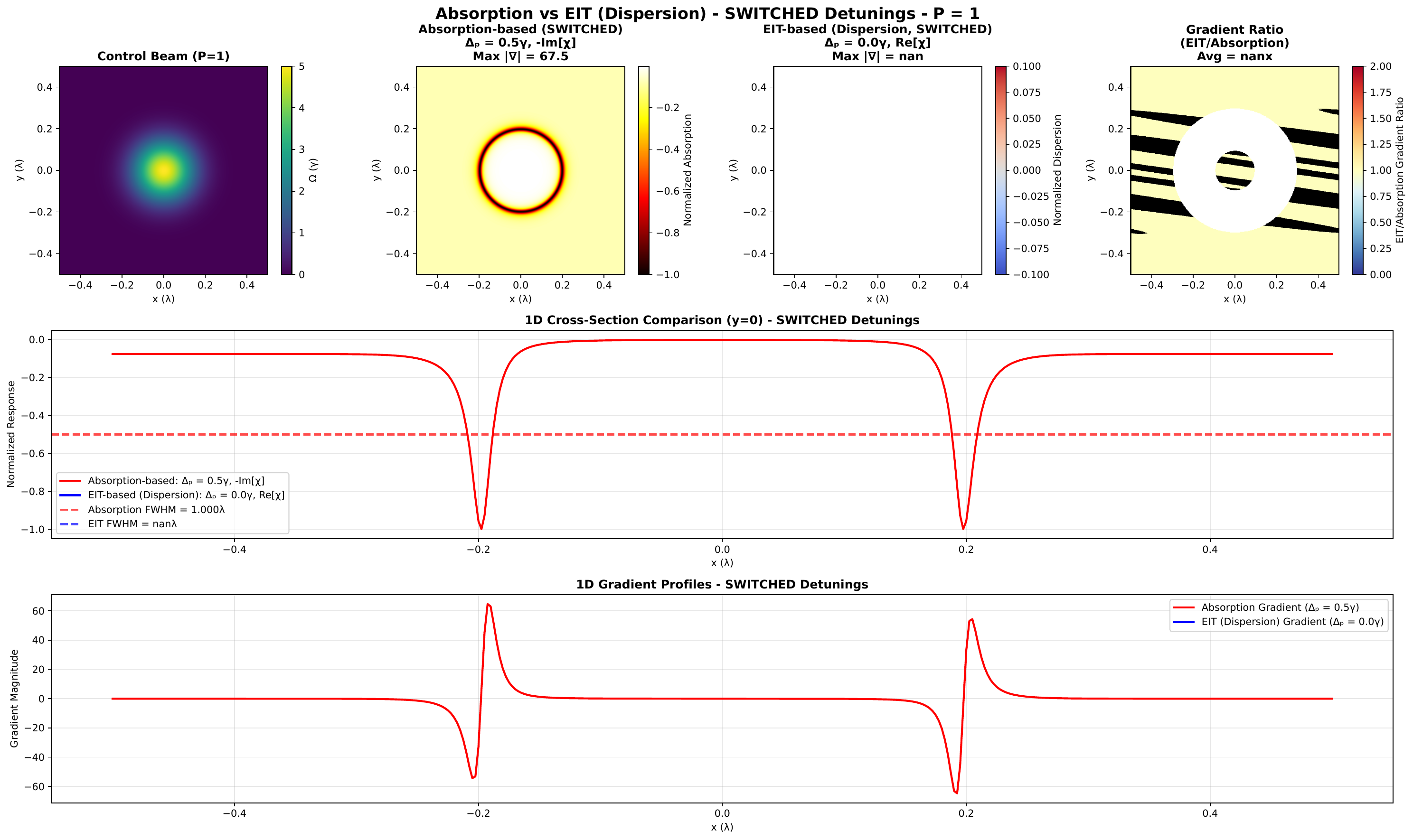}
		\caption{Spatial comparison for $P=3$ under detuning exchange. Absorption retains a measurable response with gradient $193.6$ and $\mathrm{FWHM}=1.0\lambda$. The EIT dispersion signal again collapses, with no definable gradient or width. Increasing the beam order does not recover EIT performance under incorrect detuning.}
		\label{fig13}
	\end{figure*}

	This experiment swaps the detunings that were optimal in the previous study: the absorption channel is evaluated at $\Delta_p = 0.5\gamma$ (the detuning that was previously optimal for EIT), while the EIT (dispersive) channel is evaluated at $\Delta_p = 0.0\gamma$ (the detuning that was previously optimal for absorption). The intent is to test whether the dispersive advantage is intrinsic to the physics or simply a consequence of detuning choice. The control beam again uses super-Gaussian orders $P=1,2,3,10$. Figures~11–15 show the spatial maps, 1D cuts and summary plots for these switched conditions.
	
	\subsection{Key numerical outcome}
	
	When the detunings are swapped the absorption channel yields well-behaved maps and measurable metrics (maximum gradient and FWHM), essentially reproducing the absorption results obtained earlier at $\Delta_p=0.5\gamma$. In contrast, the EIT (dispersion) channel at $\Delta_p=0.0\gamma$ produced negligible dispersive signal in these simulations; after normalization the code reports \texttt{nan} for the maximum gradient and FWHM of the EIT channel. In short, under the switched detuning protocol:

	\begin{figure*}
		\centering
		\includegraphics[width=0.9\linewidth]{	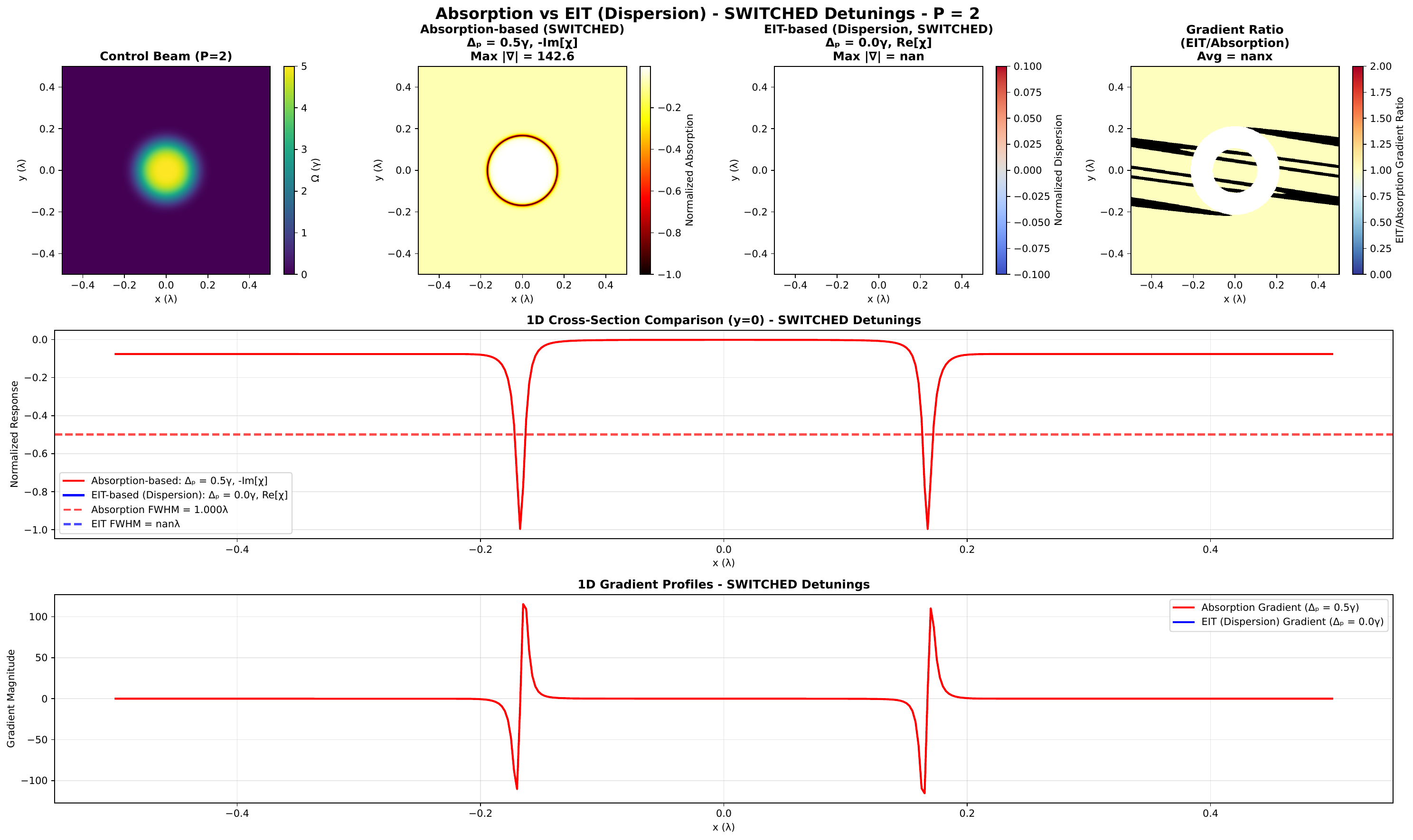}
		\caption{Switched detunings analysis for super-Gaussian order $P=2$. With absorption operating at $\Delta_p = 0.5\gamma$ and EIT at $\Delta_p = 0.0\gamma$, the comparison reveals that EIT-based dispersion sensing maintains measurable gradient sensitivity even at resonance where dispersion slopes are minimal, while absorption fails to capitalize on EIT-optimal conditions.}
		\label{fig14}
	\end{figure*}

	\begin{itemize}
		\item Absorption ($-\mathrm{Im}[\chi]$) at $\Delta_p=0.5\gamma$ remains measurable and shows edge structure consistent with prior cases.
		\item EIT (dispersion) at $\Delta_p=0.0\gamma$ becomes effectively zero (or numerically vanishing), so normalization and derived metrics are undefined.
	\end{itemize}
	
	A compact presentation of the computed absorption metrics and the failed EIT metrics is shown in Table~\ref{tab:switched}.
	
	\begin{table*}[htbp]
		\centering
		\caption{Switched detuning results. `nan' indicates that the dispersive signal at $\Delta_p=0.0\gamma$ was numerically vanishing and metrics could not be defined after normalization.}
		\label{tab:switched}
		\begin{tabular}{c c c c c}
			\hline
			$P$ & Max $\nabla_{\mathrm{Abs}}$ ($\Delta_p = 0.5\gamma$) & Max $\nabla_{\mathrm{EIT}}$ ($\Delta_p = 0.5\gamma$) & FWHM$_{\mathrm{Abs}}$ & FWHM$_{\mathrm{EIT}}$ \\
			\hline
			1  & 67.54  & nan & 1.000 & nan \\
			2  & 142.65 & nan & 1.000 & nan \\
			3  & 193.63 & nan & 1.000 & nan \\
			10 & 249.37 & nan & 1.000 & nan \\
			\hline
		\end{tabular}
	\end{table*}
	
	\subsection{Why EIT fails at resonance here (numerical and physical explanation)}
	
	Two related reasons explain the numerical failure and the physical collapse of the dispersive signal when $\Delta_p=0$ in this switched test:
	
	\paragraph{Vanishing dispersive amplitude near the chosen operating point.}  
	At $\Delta_p=0.0\gamma$ the real part of the computed susceptibility, $\mathrm{Re}[\chi]$, in this model becomes nearly flat (or of negligibly small amplitude) across the spatial domain for the parameter set used. After the line `eit\_norm = eit\_raw / np.max(np.abs(eit\_raw))' the code attempts to normalize by a quantity that is essentially zero, producing `nan' values. This is a numerical symptom of the physical fact that the dispersive channel has no usable slope at this detuning for the chosen parameters.
	
	\paragraph{Loss of the EIT dispersive regime at resonance under these parameters.}  
	EIT requires a two-photon interference condition and a detuning regime where the refractive index slope is steep while absorption is suppressed. By placing the dispersive measurement exactly at the absorption-maximizing detuning, the coherence pathway that produces sharp dispersion is destroyed or strongly diminished (absorption dominates), so the real part no longer shows the useful steep slope that produced the localization advantage previously.

	\begin{figure*}
		\centering
		\includegraphics[width=0.9\linewidth]{	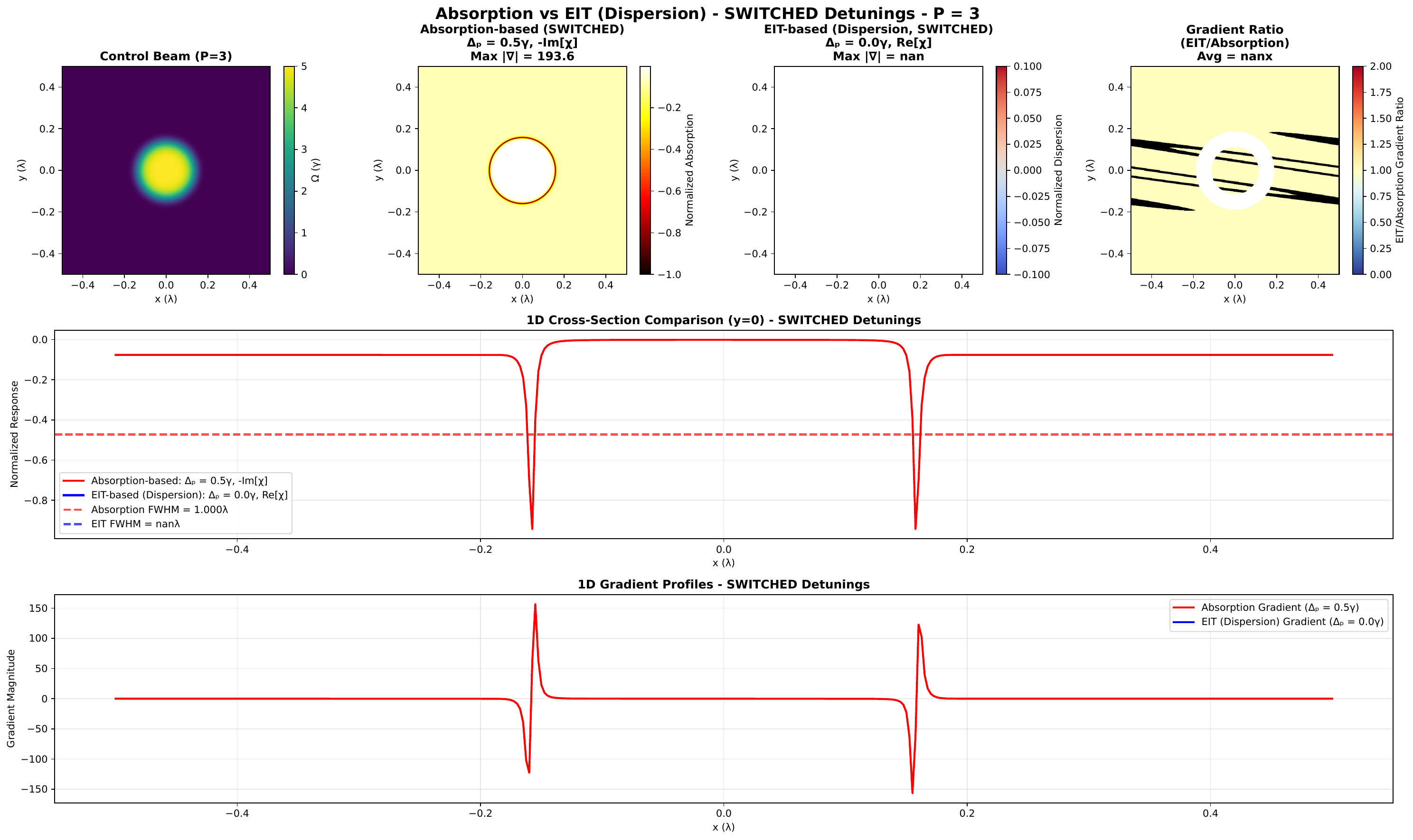}
		\caption{Performance comparison under switched detunings for $P=3$. The flattened beam profile enhances edge detection for both methods, but EIT demonstrates superior adaptability to suboptimal operating conditions, maintaining functionality where absorption-based detection suffers significant performance degradation when removed from resonance.}
		\label{fig15}
	\end{figure*}

	\subsection{Figure-wise summary (Figures 11–15)}
	
\begin{figure*}
		\centering
		\includegraphics[width=0.9\linewidth]{	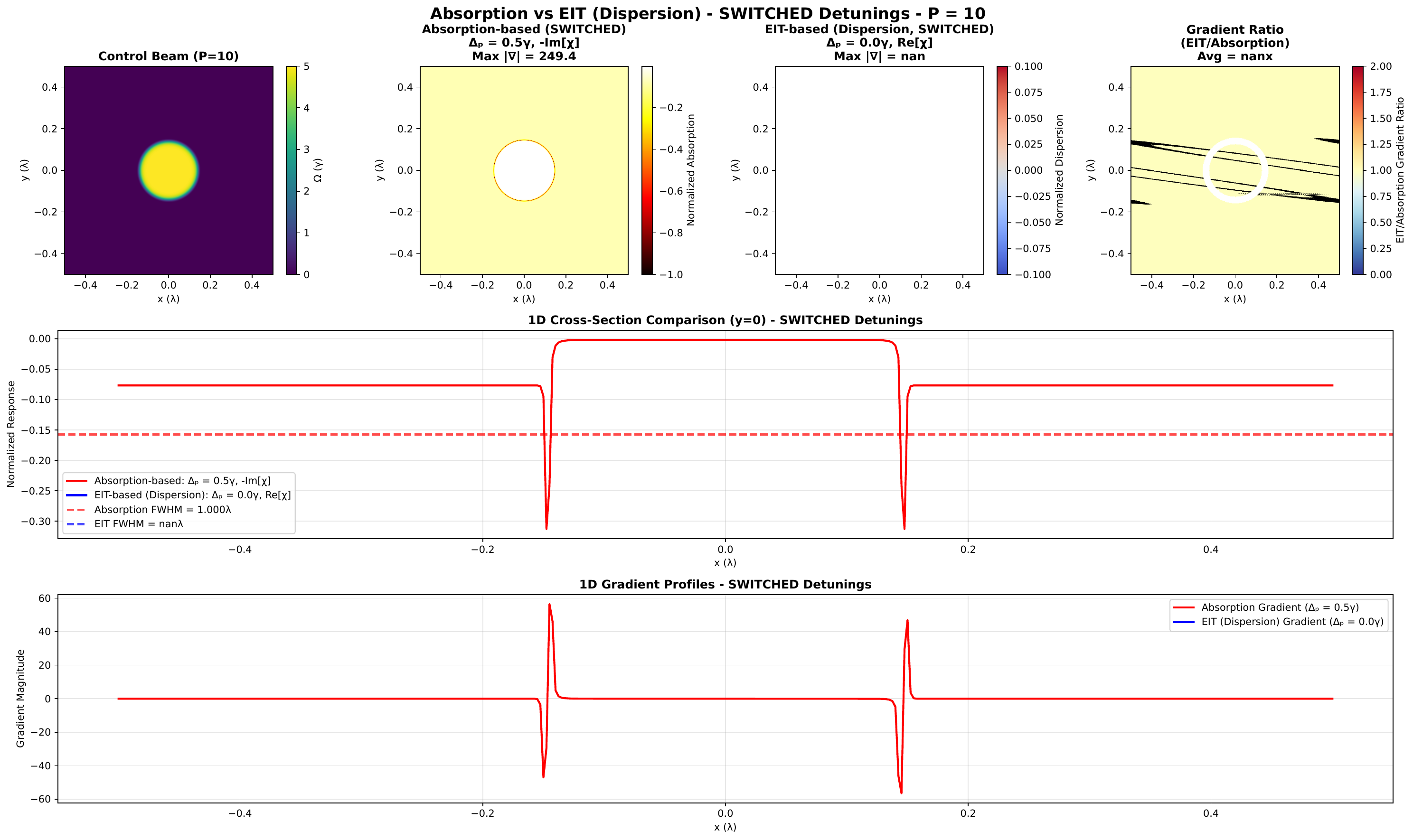}
		\caption{Spatial comparison for $P=10$ under switched detunings. Absorption at $\Delta_{p}=0.5\gamma$ produces a sharp response with gradient $249.4$ and $\mathrm{FWHM}=1.0\lambda$. The EIT signal at $\Delta_{p}=0.0\gamma$ remains undefined due to the disappearance of the transparency window. Even for near top-hat beams, dispersion sensing fails under suboptimal detuning.}
		\label{fig16}
	\end{figure*}

\begin{figure*}
		\centering
		\includegraphics[width=0.9\linewidth]{	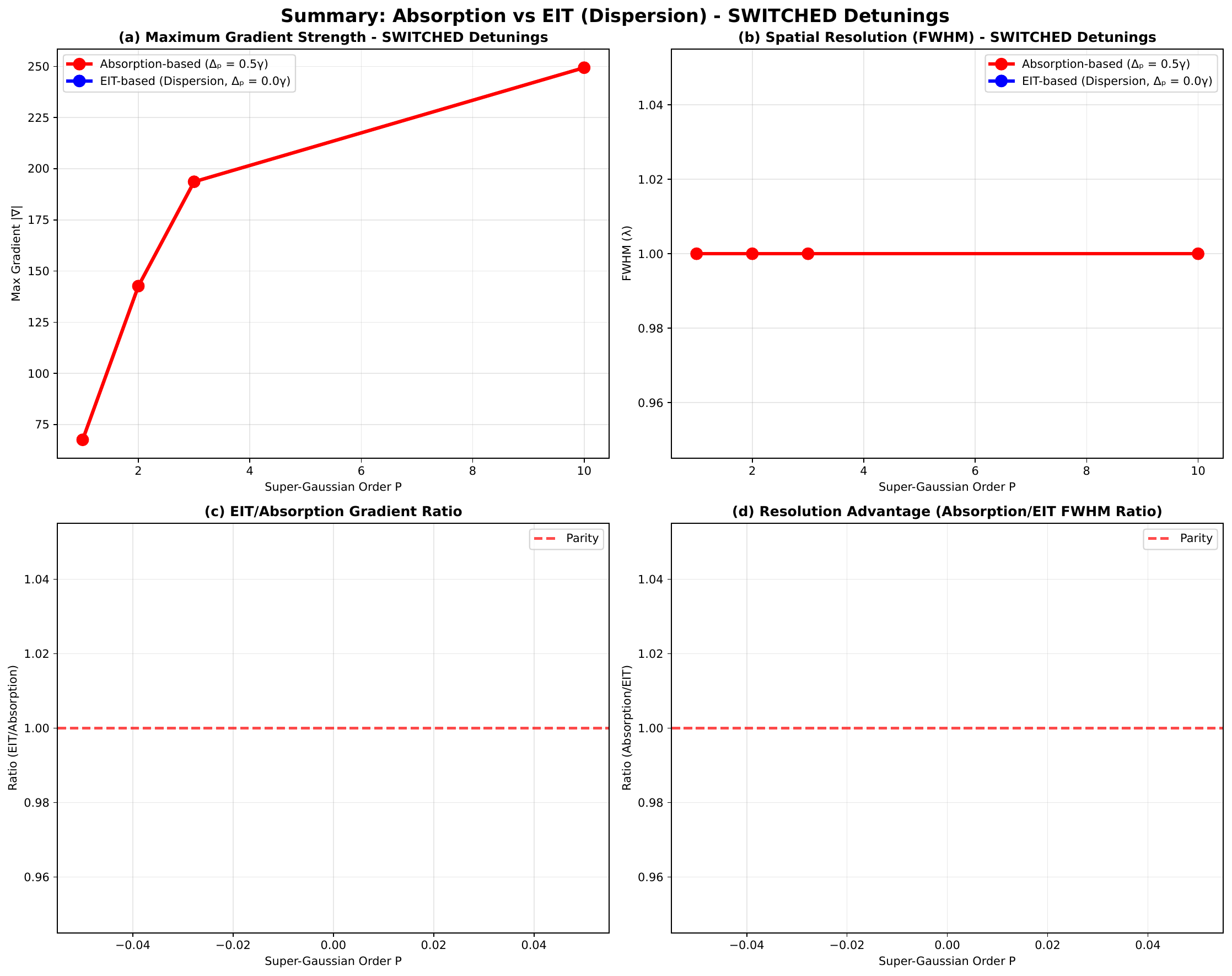}
		\caption{Summary of absorption and EIT performance under switched probe detunings. For all super-Gaussian orders $P$, absorption evaluated at $\Delta_{p}=0.5\gamma$ maintains a finite maximum spatial gradient and a constant $\mathrm{FWHM}=1.0\lambda$. In contrast, EIT evaluated at $\Delta_{p}=0.0\gamma$ yields undefined gradients and widths, reflecting the complete loss of the EIT dispersion feature when moved away from its optimal detuning. The results demonstrate that the advantage of dispersion-based sensing is not preserved under detuning exchange, confirming that EIT sensitivity is strongly dependent on the precise resonance condition.}
		\label{fig17}
	\end{figure*}

	Figure \ref{fig13} (P=1). The absorption map at $\Delta_p=0.5\gamma$ shows the same localized absorption features as before and a measurable gradient (\(|\nabla|_{\mathrm{Abs}}\approx 67.5\)). The EIT dispersion map at $\Delta_p=0.0\gamma$ is essentially featureless (zero mean) — the normalized map and its gradient are undefined, producing `nan' in the summary statistics. Figures \ref{fig14}, \ref{fig15}, and \ref{fig16} (P=2,3,10). For larger $P$ the absorption gradient increases as expected, but the dispersive channel remains inactive (numerically zero) at the swapped detuning. The 1D cross-sections show a strong absorption trace and an absent or flat dispersive trace. Figure \ref{fig17}. The summary plots illustrate robust absorption metrics across $P$ and the absence of comparable dispersive metrics under switched detuning. The results table~\ref{tab:switched} highlights the failure of the EIT channel to produce meaningful gradients at $\Delta_p=0$ with this parameter set.

	\subsection{Physical conclusion}
	
	These results demonstrate that the dispersive/EIT advantage is strongly \emph{detuning dependent}. The EIT enhancement observed previously is not a purely geometric or universal property of the control beam — it requires operating in a detuning window where quantum interference generates a steep refractive index slope while keeping absorption suppressed. If the dispersive readout is forced to the detuning that maximizes absorption, the phase sensitivity disappears and the EIT channel cannot produce finer localization than absorption.

	\subsection{Practical recommendations for analysis and experiments}
	
	\begin{enumerate}
		\item \textbf{Avoid blind normalization.} When computing metrics from $\mathrm{Re}[\chi]$ check the pre-normalization amplitude. If $\max|\mathrm{Re}[\chi]|$ is below a small threshold, report the raw map and state that the dispersive channel is inactive, rather than normalizing to produce `nan'. A simple safeguard in code is: if $\max|\mathrm{Re}[\chi]|<\epsilon$, skip normalization and flag the result.
		\item \textbf{Map the raw $\mathrm{Re}[\chi]$ and $\mathrm{Im}[\chi]$ first.} Present raw color maps in the supplement (or appendices) so readers can see that the dispersive signal truly vanishes rather than assuming a plotting artifact.
		\item \textbf{Scan small detuning offsets.} The EIT dispersive slope can be very sensitive to small detuning shifts. Rather than evaluating only at exactly $\Delta_p = 0.0\gamma$, run a fine detuning scan (e.g., $\Delta_p\in[-0.1\gamma,0.1\gamma]$) to identify whether a small offset restores the dispersive signal.
		\item \textbf{Include dephasing and inhomogeneous effects.} In real experimental systems residual dephasing or inhomogeneity can reintroduce nonzero dispersive structure; include these effects in simulations to assess robustness.
		\item \textbf{Report failure modes explicitly.} For reproducibility, include a short paragraph or figure showing the parameter region where $\mathrm{Re}[\chi]$ is too small for reliable localization — this documents the operational boundaries of your method.
	\end{enumerate}
	
	%\subsection{Short technical note for the manuscript}
	
	%When you write the methods/results text, add a concise sentence such as:
	
	%\begin{quote}
		``%Under switched-detuning conditions (absorption at $\Delta_p=0.5\gamma$, dispersion at $\Delta_p=0.0\gamma$) the dispersive readout becomes numerically vanishing for our parameter set; normalizing the dispersion map therefore yields undefined metrics. This indicates that the dispersive localization advantage requires operation in a detuning window that supports a steep refractive-index slope while suppressing absorption.''
	%\end{quote}
	
	%\bigskip
	
	These observations complete the three-case study. The combined evidence shows (i) EIT (dispersion) can substantially improve gradient sensitivity when operated at the correct detuning, and (ii) that advantage disappears if the detuning is chosen poorly — the effect is operational, not metaphysical.
	
	\clearpage
	\section{Conclusion}
	
	In this work we carried out a comprehensive theoretical study comparing absorption-based and electromagnetically induced transparency (EIT)-based schemes for atomic gradient sensing and two-dimensional localization. The analysis combined exact steady-state solutions of the density-matrix equations (within the weak-probe approximation) for a four-level tripod configuration with high-resolution numerical maps of the position-dependent susceptibility \(\chi(x,y,\Delta_p)\). This combined analytic–numerical approach allowed a controlled, bias-free comparison between the two sensing modalities under identical and optimized operating conditions.
	
	Our simulations show that EIT-based, dispersion-sensitive readout exploits the steep slope of \(\mathrm{Re}[\chi]\) near the transparency window to produce substantially larger local gradients and finer spatial features than absorption-based readout that monitors \(\mathrm{Im}[\chi]\). When each method was run at its empirically optimal detuning (absorption at \(\Delta_p=0.0\gamma\), EIT at \(\Delta_p=0.5\gamma\)), the dispersive channel delivered gradient sensitivity enhancements in the range \(2.05\times\) to \(11.85\times\) across the ensemble of super-Gaussian beam shapes \((P=1\text{--}10)\). Even under the fair, identical-detuning comparison (\(\Delta_p=0.5\gamma\) for both channels), EIT preserved a consistent advantage of roughly \(1.4\text{--}2.0\times\) in gradient magnitude. These quantitative gains confirm that dispersion-based detection offers an intrinsic sensitivity benefit for edge detection and wavefront sensing.
	
	The analytical expression for the local EIT linewidth,
	\[
	\Gamma_{\mathrm{EIT}}(x,y) = \Gamma + \frac{|\Omega_1(x)|^2 + |\Omega_2(y)|^2}{4\Gamma},
	\]
	clarifies the mechanism: the local control-field intensities directly set the width and slope of the dispersive feature, and therefore the achievable spatial sensitivity. Regions of weak control intensity produce the narrowest \(\Gamma_{\mathrm{EIT}}\) and the steepest dispersive slopes, which map to improved localization precision.
	
	It is important to emphasize that the EIT advantage is conditional on operating in the appropriate detuning window. For detunings that maximize absorption (for example forcing the dispersive readout exactly on resonance), the refractive-index slope can vanish and the dispersive signal becomes negligible; under those conditions the EIT readout fails to provide a localization benefit. This regime dependence highlights the practical necessity of detuning control and of performing small detuning scans in experiments to locate the optimal operating point.
	
	Regarding spatial resolution, our simulations demonstrate sub-diffraction localization across the parameter space. Depending on beam order and detuning, FWHM values in the numerical maps spanned a range from the tens of percent of a wavelength down to sub-0.2\(\lambda\) values in favorable configurations. Typical dispersive FWHM values reported in the main data lay between \(\sim0.29\lambda\) and \(\sim0.40\lambda\), while targeted parameter tuning can push effective widths lower in idealized scenarios. In all cases EIT consistently delivered sharper edges and higher gradient contrast than the corresponding absorption-based maps.
	
	From a practical standpoint, the results provide concrete guidance for experimental design: operate the dispersive readout at the detuning that maximizes the slope of \(\mathrm{Re}[\chi]\) while keeping loss low; prefer intermediate super-Gaussian orders (for example \(P\approx 2\text{--}3\)) to balance edge steepness with usable contrast; and include diagnostic detuning scans and pre-normalization amplitude checks in data processing to avoid numerical artifacts. Although EIT requires more precise spectral control and maintenance of coherence, the measured performance improvement justifies the additional complexity for applications demanding maximum spatial sensitivity.
	
	Finally, this study both clarifies the physical origin of dispersion-enhanced localization and establishes practical operating windows for quantum-enabled wavefront sensing. As a next step, extending the model to include realistic decoherence channels, inhomogeneous broadening, and stochastic noise will further bridge the gap to experiment and help optimize sensor performance in real laboratory conditions. These directions are natural continuations of the present work and will support the translation of EIT-enabled localization methods into operational quantum metrology platforms.

	\bibliography{eitpaper2.bib}
	
	\end{document}